\documentclass[]{emulateapj}

\def\rmit#1{{\it #1}}              %% italics (RR mode, Kluwer)

\def\ie{\rmit{i.e.}}
\def\eg{\rmit{e.g.}}

\def\kms{\hbox{km$\;$s$^{-1}$}}
\def\ms{\hbox{m$\;$s$^{-1}$}}

\shorttitle{Magneto-Acoustic Waves in Sunspots}

\shortauthors{Felipe et al.}

\begin{document}

\title{Magneto-acoustic waves in sunspots:  first results from a new 3D nonlinear magnetohydrodynamic code}

\author{T. Felipe\altaffilmark{1,2}, E. Khomenko\altaffilmark{1,2,3} and M. Collados\altaffilmark{1,2}}
\email{tobias@iac.es}

\altaffiltext{1}{Instituto de Astrof\'{\i}sica de Canarias, 38205,
C/ V\'{\i}a L{\'a}ctea, s/n, La Laguna, Tenerife, Spain}
\altaffiltext{2}{Departamento de Astrof\'{\i}sica, Universidad de La Laguna, 38205, La Laguna, Tenerife, Spain}
\altaffiltext{3}{Main Astronomical Observatory, NAS, 03680, Kyiv,
Ukraine}

\begin{abstract}
Waves observed in the photosphere and chromosphere of sunspots
show complex dynamics and spatial patterns. The interpretation of
high-resolution sunspot wave observations requires modeling of three-dimensional non-linear wave propagation and mode
transformation in the sunspot upper layers in realistic spot
model atmospheres. Here we present the first results of such modeling. We
have developed a 3D non-linear numerical code specially designed
to calculate the response of magnetic structures in
equilibrium to an arbitrary perturbation. The code solves the 3D
nonlinear MHD equations for perturbations; it is stabilized by
hyper-diffusivity terms and is fully parallelized. The robustness
of the code is demonstrated by a number of standard tests. We
analyze several simulations of a sunspot perturbed by pulses of
different periods at subphotospheric level, from short periods,
introduced for academic purposes, to longer and realistic periods
of three and five minutes. We present a detailed description of the
three-dimensional mode transformation in a non-trivial sunspot-like
magnetic field configuration, including the conversion between
fast and slow magneto-acoustic waves and the Alfv\'en wave, by
calculation of the wave energy fluxes. Our main findings are the
following: (1) the conversion from acoustic to the Alfv\'en mode
is only observed if the the driving pulse is located out of the
sunspot axis, but this conversion is energetically inefficient;
(2) as a consequence of the cut-off effects and refraction of the
fast magneto-acoustic mode, the energy of the evanescent waves
with periods around 5 minutes remains almost completely below the
level $\beta=1$; (3) waves with frequencies above the
cut-off propagate field-aligned to the chromosphere and their
power becomes dominating over that of evanescent 5-minute
oscillations, in agreement with observations.
\end{abstract}


%%%%%%%%%%%%%%%%%%%%%%%%%%%%%%%%%%%%%%%%%%%%%%%%%%%%%%%%%%%%%%%%%%%%%%%%%%%

\section{Introduction}

Sunspots can be considered as laboratories for studies of
magnetized plasma in conditions that are inaccessible on Earth.
They give clues about the physics of energy propagation (\eg, in
the form of different oscillatory modes) in fluids permeated by
strong magnetic fields. From the analysis of wave propagation it
is possible to infer basic properties of stellar atmospheres.
Waves observed in active regions are believed to have a relevant
role in the energy balance of the solar atmosphere, being one of
the candidates to explain the chromospheric heating. The first
detection of waves in sunspots was done by
\citet{Beckers+Tallant1969}, who called them ``umbral flashes''.
%It was done seven years after the first measurements of
%oscillations in quiet Sun by \citet{Leighton+etal1962}.
Since this
date waves have been found in many solar features, from photospheric flux tubes to the solar wind.

Oscillations show a different behavior in different regions of
sunspots. In the umbral photosphere their power spectrum peaks
around 3 mHz (period of 5 minutes), and their amplitudes are around
a hundred \ms. Waves in the photosphere of sunspot umbrae are
similar to those observed in the quiet sun, but with their power
reduced by a factor of 2--5 \citep{Woods+Cram1981}. Several
mechanisms have been proposed to explain this power suppression:
reduction of wave excitation inside sunspots
\citep{Goldreich+Keeley1977, Goldreich+Kumar1988,
Goldreich+Kumar1990}, $p$-mode absorption inside sunspots
\citep{Cally1995}, different heights of spectral line formation
due to the Wilson depression and altering of $p$-mode
eigenfunctions by the magnetic field \citep{Hindman+etal1997}.

Chromospheric umbral oscillations have a major power peak around 5
mHz (period of 3 minutes) and amplitudes of several kilometers per
second. Velocities measured in chromospheric umbrae show saw-tooth
temporal profiles, typical for shock waves \citep{Lites1986,
Centeno+etal2006a}. To explain the multiple peaks in the 3 minute
band in the power spectrum of the chromospheric waves several
hypotheses have been investigated: a resonant chromospheric cavity
\citep{Zhugzhda+etal1985}; non-linear interaction of photospheric
modes \citep{Gurman+Leibacher1984}; slow magneto-acoustic mode
field-aligned propagation from the photosphere to the chromosphere
in the 5--6 mHz band \citep{Centeno+etal2006a}.

In the penumbra the dominant oscillatory phenomena are running
penumbral waves, which are best seen in H$\alpha$ core velocity
observations as disturbances propagating radially outward from the
umbra. As they move across the penumbra their radial velocity is
apparently reduced and their frequencies also decrease from $4-5$
mHz near the umbral/penumbral boundary to around $0.7-1.5$ mHz at
the outer edge of the penumbra \citep{Bogdan+Judge2006}.

It is now becoming apparent that the photospheric 5-minute
oscillations, chromospheric 3-minute oscillations, and running
penumbral waves are the different manifestation of the same global
dynamical phenomenon \citep{Rouppe+etal2003,
Bloomfield+etal2007b}. No fully comprehensive model of this global
oscillation phenomenon was presented up to date due to the
complicated mathematical description of the physical processes
playing a role in realistic magneto-atmospheres. The recent
progress in observations and numerical simulations of sunspot
waves is summarized in \citet{Khomenko2010}.

Although a lot of analytical work was done in simple atmospheres
\citep[][to name a few]{Ferraro+Plumpton1958,
Zhugzhda+Dzhalilov1984a}, those works were restricted to very
idealized cases. Numerical simulations allow more flexibility and
over the last years many attempts were done to perform numerical
modelling of waves in non-trivial magnetic field configurations,
with applications to the photosphere and the chromosphere
\citep[\eg][]{Cargill+etal1997, Rosenthal+etal2002,
Hasan+etal2003, Bogdan+etal2003, Hasan+Ulmschneider2004,
Hasan+etal2005, Khomenko+Collados2006, Khomenko+etal2008a}. In
most of the cases, these works were restricted to studies of the
behavior of high-frequency waves (above cut-off) in
two-dimensional situations. Despite these limitations, several
important questions were learned from these models. It was shown
that the fast magneto-acoustic mode in the magnetically dominated
region (\ie\ where the sound speed $c_S$ is much lower than the
Alfv\'en speed $v_A$) is refracted and reflected down to the gas
pressure dominated atmosphere due to the gradients of the Alfv\'en
speed \citep{Khomenko+Collados2006}. Earlier,
\citet{Rosenthal+etal2002} demonstrated that the inclination of
the magnetic field lines is important for the fast mode
reflection, \ie\ in those regions where the inclination angle is
large almost all of the fast mode wave energy is reflected back
down. Another important feature of all these simulations is the
presence of the mode transformation at the layer where $v_A=c_S$.
Around this layer, the phase speeds of all modes are similar and
different waves can interact \citep{Bogdan+etal2003, Cally2006,
Khomenko+Collados2006}. The direction and efficiency of the mode
transformation depends on the frequency of the wave and the angle
between the wave vector and the magnetic field \citep{Cally2005,
Cally2006}. When this angle is small and the frequency is high, the
fast mode can be converted into the slow mode and vice versa.

Most of the works cited above on the simulations of wave
propagation in the upper layers of sunspots, as well as the
analytical theories of the mode transformation, were developed for
high-frequency waves with frequencies above, or just at, the
cut-off frequency. There is another class of simulations where the
problem of helioseismic wave propagation below sunspots is
addressed and waves with realistic solar frequencies in the 3-5
mHz range are considered \citep{Cally+Bogdan1997,
Parchevsky+Kosovichev2009, Hanasoge2008, Cameron+etal2008,
Moradi+etal2009, Khomenko+etal2009}. In the latter works most
attention has been paid to the wave propagation in sub-surface
layers, rather than in upper photospheric and chromospheric layers.
Thus, there is gap between these two kind of models. With the
development of our code and its first results presented here we
pretend to cover this gap and to address the problem of three
dimensional propagation and transformation of the 3-5 mHz period
waves in the upper layers of sunspots. Our aim is to investigate
the response of the magnetic atmospheres to oscillations with low
frequencies (like 5-minute waves) and to compare this response
with that from the high-frequency perturbations. As those 5-minute
waves have most power at the photospheric heights and are used in
measurements to derive the parameters of the solar atmosphere,
including solar active regions, understanding their propagation
and transformation properties is important. Our objective is to
perform numerical calculations sufficiently realistic to imitate
the wave excitation in sunspots, reproduce the change of wave
frequency with height, formation of shocks at chromospheric
layers, and to be at the level allowing comparison with
photospheric and chromospheric observations by spectral synthesis.

Several numerical codes have been developed among the solar
physics community to study the propagation of waves in magnetized
atmospheres. They adopt different strategies for the numerical
scheme, boundary conditions and wave driving, having their
advantages and disadvantages \citep{Moradi+etal2010}. The upper
boundary typically represents a problem in wave simulations, since
waves should not artificially be reflected there back into the
physical domain. \citet{Rosenthal+etal2002} apply characteristic
boundary conditions at the top boundary; \citet{Hasan+etal2005}
uses the open boundary concept; while \citet{Khomenko+etal2008a}
introduced a special media at the top called Perfectly Matched
Layer (PML) which absorbs with almost no reflections the waves
that reach the upper boundary. \citet{Rosenthal+etal2002} and
\citet{Hasan+etal2005} solve the complete MHD equations, while
\citet{Khomenko+etal2008a} solve equations for perturbations with
all non-linear terms retained. This strategy gives them an
advantage for precision of the numerical scheme and for the
application of the boundary conditions.

3D MHD codes for wave simulations are starting to be available.
\citet{Cameron+etal2007} presented the semi-spectral linear MHD
code SLiM, developed for helioseismology purposes. In this code,
the horizontal derivatives are evaluated in Fourier space and the
upper boundary is treated as a sponge layer. Another 3D linear MHD
code for wave propagation has been developed by
\citet{Parchevsky+Kosovichev2007}. The authors use the realistic
OPAL equation of state and a PML layer as the upper boundary
condition. Numerically, the upper magnetized atmospheric layers
represent an additional problem limiting strongly the time step of
the simulations due to the high values of the Alfv\'en speed. To
overcome this problem, one of the strategies used is the Lorentz
force controller. This method consists in reducing the amplitude
of the Lorentz force in the layers where the Alfv\'en speed is
large. This method is used by \citet{Hanasoge2008} in his 3D
linear code. However, the influence of this artificial procedure
on the simulated wave properties has not been verified yet.
Recently \citet{Shelyag+etal2008} presented a nonlinear 3D
parallel code developed on the base of the VAC code
\citep{Toth1996}. In this code, they use the same philosophy as
\citet{Khomenko+etal2008a}, solving non-linear equations for
perturbations. As for sunspots, only results of the 2D
helioseismic wave propagation below surface have been presented so
far from this code \citep{Shelyag+etal2009}, not addressing the
wave propagation in the upper layers.

The non-linear MHD code described in this paper represents an
extension of the code by \citet{Khomenko+etal2008a} into three
dimensions. Our code solves the MHD equations for perturbations in
a background plasma in magnetohydrostatic equilibrium using
artificial diffusion to assure numerical stability of the
solution, and is fully MPI-parallelized. In this paper we present
extensive numerical tests demonstrating the code stability and
robustness. We also present the first scientific results of the
three-dimensional wave propagation and transformations in the
upper atmosphere of a sunspot model. Our first aims are to study
the propagation of waves, excited by sub-photospheric pulses, up
to the chromosphere, including the shock wave formation. We
present a detailed description of the three-dimensional mode
transformation for waves with different frequencies, from below to
above the acoustic cut-off frequency. In particular, we address
the question of conversion to Alfv\'en waves in realistic
conditions. Works on conversion from/to Alfv\'en waves in three
dimensions are only now being initiated
\citep[\eg][]{Cally+Goossens2008} and no generalized picture has
been obtained yet. The organization of the paper is as follows. In
Section \ref{sect:numericalprocedure} we describe the MHD
equations and the numerical details of the calculations. In
Section \ref{sect:analisis} we discuss the results of the
simulations of the three-dimensional wave propagation and mode
transformation in a sunspot model. This section includes the
description of the magneto-static sunspot model in equilibrium
(Section \ref{sect:MHS}) and a brief discussion of the dispersion
relations and wave propagation speeds (Section
\ref{sect:dispersion}). Section \ref{sect:conclusiones} presents
our conclusions and our future plans for the application of the
code. Finally, in Appendix \ref{sect:test},
we show several numerical tests applied to prove the robustness
of the code.

\section{Numerical procedure}
\label{sect:numericalprocedure}

\subsection{MHD equations}

Our numerical code is an extension to 3 dimensions of the code
described in \citet{Khomenko+Collados2006, Khomenko+etal2008a}. It
solves the non-linear equations of ideal compressible MHD. Written
in conservative form, these equations are:

% MHD ideal equations

\begin{equation}
 \frac{\partial\rho}{\partial t}+\nabla(\rho{\bf v})=0 \,, \label{eq:den}
\end{equation}
\begin{equation}
\frac{\partial (\rho{\bf v})}{\partial t}+\nabla\cdot\Big
[\rho{\bf vv}+\Big (p+\frac{{\bf B}^2}{2 \mu_0}\Big ){\bf
I}-\frac{{\bf B}{\bf B}}{\mu_0}\Big ]=\rho{\bf g}+{\bf S}(t) \,,
\label{eq:mom}
\end{equation}
\begin{equation}
\frac{\partial e}{\partial t}+\nabla\cdot\Big[{\bf
v}\Big(e+p+\frac{|{\bf B}|^2}{2 \mu_0}\Big)-\frac{1}{\mu_0}{\bf
B}({\bf v\cdot B})\Big ]=\rho ({\bf g\cdot v})+Q_{\rm rad} \,,
\label{eq:ene}
\end{equation}
\begin{equation}
 \label{eq:ind}
\frac{\partial {\bf B}}{\partial t}=\nabla\times ({\bf v} \times
{\bf B}) \,,
 \end{equation}

\noindent where ${\bf I}$ is the identity tensor, $\rho$ is the
density, ${\bf v}$ is the velocity, $p$ is the gas pressure, ${\bf
B}$ is the magnetic field, ${\bf g}$ is the gravitational
acceleration and $e$ is the total energy per unit volume,
\begin{equation}
e=\frac{1}{2}\rho v^2 + \frac{p}{\gamma-1} + \frac{B^2}{2\mu_0} \,.
 \end{equation}

The dot '$\cdot$' represents the scalar product of vectors, while
the notation '${\bf B}{\bf B}$' stands for the tensor product. The
energy losses $Q_{\rm rad}$ can be described by the Newton law of
cooling, but, in the simulations presented in this paper, it is
set to zero. We neglect the viscous force, the thermal conduction
and the terms describing the diffusion of the magnetic field.
However, artificial equivalents of some of these terms are
introduced later for the issue of numerical stability of the
simulations. The term {\bf S}(t) in Eq. \ref{eq:mom} represents a
time-dependent external force.

%MHS equilibrium
In an equilibrium state, where temporal derivatives and velocities
are null, and in the absence of external forces ($\textbf{S}=0$),
the previous equations reduce to the equations of the force
balance for a gravitationally stratified magnetized plasma:

\begin{equation}
\nabla\cdot\Big [\Big (p_0+\frac{{{\bf
B}_0}^2}{2 \mu_0}\Big ){\bf I}-\frac{{\bf B}_0{\bf B}_0}{\mu_0}\Big ]=\rho_0{\bf g}\,.
\label{eq:mhs}
\end{equation}

%Perturbed variables

Considering departures from the equilibrium state induced by an
external force {\bf S}, $\rho$, $p$ and ${\bf B}$ can be expressed
as the sum of the background value (subindex 0) and the
perturbation (subindex 1):

\begin{equation}
\rho=\rho_0+\rho_1\,,
\label{eq:rho1}
\end{equation}

\begin{equation}
p=p_0+p_1\,,
\label{eq:p1}
\end{equation}
and
\begin{equation}
{\bf B}={\bf B_0}+{\bf B}_1\,,
\label{eq:b1}
\end{equation}

\noindent while the velocity only corresponds to a perturbed value ${\bf v}={\bf v}_1$.

%MHD equations for perturbation
The non-linear equations for perturbations are obtained by
replacing expressions (\ref{eq:rho1}--\ref{eq:b1}) in
Eqs.~(\ref{eq:den}--\ref{eq:ind}) and subtracting the equation of
the magnetohydrostatic equilibrium (Eq.~\ref{eq:mhs}). The
following system of MHD equations for perturbations of density,
pressure, magnetic field and velocities is obtained in
conservative form:

\begin{equation}
\label{eq:den1} \frac{\partial\rho_1}{\partial
t}+\nabla\Big[(\rho_0+\rho_1){\bf v_1}\Big]= 0 \,,
\end{equation}

\begin{eqnarray}
\lefteqn{\frac{\partial [(\rho_0+\rho_1){\bf v}_1]}{\partial t}+\nabla \cdot\Big [(\rho_0+\rho_1){\bf
v}_1{\bf v}_1+\Big(p_1+}\nonumber\\
&&+\frac{{\bf B}_1^2}{2\mu_0}+\frac{{\bf B}_1{\bf
B}_0}{\mu_0}\Big){\bf I}- \frac{1}{\mu_0}({\bf B}_0{\bf B}_1-{\bf
B}_1{\bf
B}_0-{\bf B}_1{\bf B}_1)\Big ]=\nonumber\\
&&=\rho_1 {\bf g}+\Big (\frac{\partial [(\rho_0+\rho_1){\bf v}_1]}{
\partial t} \Big)_{\rm diff}+{\bf S(t)}\,, \label{eq:mom1}
\end{eqnarray}

\begin{eqnarray}
\lefteqn{\frac{\partial e_1}{\partial t}+\nabla\cdot\Big[{\bf
v_1}\Big((e_0+e_1)+(p_0+p_1)+}\nonumber\\
&&+\frac{|{\bf B}_0+{\bf B}_1|^2}{2 \mu_0}\Big)- \frac{1}{\mu_0}({\bf B}_0+{\bf B}_1)\Big({\bf v}_1 \cdot ({\bf B}_0 + {\bf B}_1)\Big)\Big ]=\nonumber\\
&&=(\rho_0+\rho_1) ({\bf g\cdot v_1})+Q_{\rm rad} +\Big (\frac{\partial e_1}{\partial t} \Big)_{\rm diff}\,,
\label{eq:ene1}
\end{eqnarray}

\begin{equation}
\label{eq:ind1}
\frac{\partial {\bf B}_1}{\partial t}=\nabla\times [{\bf v_1} \times ({\bf B}_0+{\bf B}_1)]+ \Big (\frac{\partial{\bf
B}_1}{\partial t} \Big)_{\rm diff} \,,
\end{equation}

Artificial diffusion terms have been added to Eqs.
(\ref{eq:mom1}--\ref{eq:ind1}) compared to Eqs.
(\ref{eq:mom}--\ref{eq:ind}). The diffusivity terms in Eqs.
(\ref{eq:mom1}--\ref{eq:ind1}) have their physical counterparts
and are needed for reasons of stability of the simulations.
A similar strategy is applied in the MURAM code
\citep{Vogler+etal2005}.

The code solves the above system of non-linear equations for
perturbations. The use of equations for perturbations instead of the
complete equations for wave simulations has two big advantages.
Firstly, the terms describing the static model and those for
perturbations can vary by orders of magnitude. Thus, by excluding
equilibrium terms we avoid important numerical precision problems.
Secondly, the boundary conditions are easier to implement on
equations for perturbations (see \S\ \ref{sect:boundary}).

In high layers, where the magnetic pressure is much larger than the
gas pressure, Eq. (\ref{eq:ene1}) is numerically problematic
as recovering thermal energy ($p$) from total energy ($e$) leads
to numerical errors. In these layers it is replaced by the
equation describing the balance of the internal energy

\begin{eqnarray}
\label{eq:int_ene1} \lefteqn{\frac{\partial p_1}{\partial t}+{\bf
v_1}\nabla(p_0+p_1)+c_S^2\Big
[\nabla \Big ((\rho_0+\rho_1){\bf v_1}\Big)-}\nonumber\\
&&-{\bf v}_1\nabla (\rho_0+\rho_1) \Big]=(\gamma -1)\rho_1Q_{\rm
rad} +\Big (\frac{\partial p_1}{\partial t} \Big)_{\rm diff}\,.
\end{eqnarray}

The transition from one equation to the other is done smoothly
using the plasma parameter $\beta$ as a criterion.

The gravity {\bf g} is constant with height and the equation of
state of an ideal gas with constant $\gamma$ is used to link the thermal
variables.

%Discretizacion espacial y temporal
The computational domain is discretized using a three-dimensional
Cartesian grid with constant space step in each dimension. The
spatial derivatives are approximated by a centered, fourth order
accurate, explicit finite differences scheme using five grid
points \citep{Vogler+etal2005}. The solution is advanced in time
by an explicit fourth-order Runge-Kutta. The fourth order
differences allow to give an accurate solution, still resolving
shock fronts, which is difficult with higher-order methods.

\subsection{Artificial diffusivity}
\label{sect:diffusivity}
%Difusividad artificial
To damp high-frequency numerical noise on sub grid scales, we
replace the physical diffusive terms in the equations of momentum
and energy by artificial equivalents. In the induction equation we
retain the magnetic diffusion term, replacing $\eta$ by an
artificial value. In general, we use a philosophy similar
to \citet{Stein+Nordlund1998}, \citet{Caunt+Korpi2001} and
\citet{Vogler+etal2005}. Each physical quantity has its own
diffusivity coefficient (scalar/vectorial for scalar/vectorial
quantities), which is formed by a shock resolving term, a
hyperdiffusivity part and a constant contribution:

\begin{center}
\begin{equation}
\label{eq:nu} \nu_l(u)=\nu_l^{\rm shk}(u)+\nu_l^{\rm
hyp}(u)+\nu_l^0,
\end{equation}
\end{center}

\noindent where $\nu_l^0=(c_S+v_A)\Delta x_lF(x,y,z)$, $u$ is the
corresponding quantity, $F(x,y,z)$ gives the form of the constant
contribution and $\Delta x_l\equiv\Delta x$, $\Delta y$, $\Delta
z$.

\subsection{Time step}
%Paso temporal
The time step has to ensure that the physical dependence domain is
included inside the numerical dependence domain. According to
this, the mesh width must be bigger than the distance traveled by
the information in a single time step due to mass flow, waves or
diffusion transport. The time step must be chosen to be smaller
than the advective time step and the time step imposed by the
diffusion terms:

\begin{equation}
\label{eq:tmin} \Delta t\leq \min(\Delta t_{\rm v},\Delta t_{\rm
diff})
\end{equation}

\noindent In this expression $\Delta t_{\rm v}$ is the time step
imposed by a modified CFL condition, approximately valid for MHD
equations,

\begin{equation}
\label{d}
\Delta t_{\rm v}=\Big [\frac{c_{\rm v}}{1/\Delta x^2+1/\Delta
y^2+1/\Delta z^2}\Big ]^{1/2}\frac{1}{v_{\rm max}}
\end{equation}

\noindent where $v_{\rm max}$ is the maximum value of the sound
and Alfv\'en speeds. The time step imposed by the diffusion $\Delta
t_{\rm diff}$ corresponds to the minimum of the diffusion time
across the three dimensions,

\begin{equation}
\label{eq:tdif} \Delta t_{\rm diff}=c_{\rm diff}
\min\Big(\frac{\Delta x^2}{\nu_x},\frac{\Delta
y^2}{\nu_y},\frac{\Delta z^2}{\nu_z} \Big )
\end{equation}

\noindent where the constant coefficients $c_{\rm v}$ and $c_{\rm diff}$
are taken to be below one to ensure the stability of the solution.
The diffusion coefficients $\nu_{x,y,z}$ are those defined in
\S\ \ref{sect:diffusivity}.

\subsection{Filtering}
%Filtrado
In the particular case of wave simulations high diffusion is not
desirable since it modifies the wave amplitudes. At the same time,
low diffusion can not always prevent the developing of high
frequency noise. In such cases we perform an additional filtering
of small wavelengths. Following \citet{Parchevsky+Kosovichev2007}
we use a sixth-order digital filter to eliminate unresolved
short-wave components:

\begin{equation}
\label{eq:filter}
u_{\rm filt}=u(x)- \sum_{m=-3}^3d_{\rm m} u(x+m\Delta x),
\end{equation}

\noindent where $u$ is a variable before filtering and $u_{\rm
filt}$ is after filtering. The filter can be applied in the three
spatial coordinates independently. The coefficients $d_{\rm m}$
have been chosen to construct the filtering function:

\begin{equation}
\label{eq:filterG} G(k\Delta x)=1- \sum_{m=-3}^3 d_m e^{imk\Delta
x}=1- \sin^6\Big (\frac{k\Delta x}{2}\Big ).
\end{equation}

\noindent The frequency of application of the filter depends on
the simulation, but it is usually applied every 10 seconds.

\subsection{Boundary conditions}
\label{sect:boundary}
%Contorno PML
Boundary conditions are an important issue for wave simulations.
One usually wants to prevent spurious wave reflections at the
boundaries. Two strategies commonly applied are based on
characteristic boundary conditions or sponge layer. Calculating
characteristic conditions \citep{Rosenthal+etal2002}, apart from
tricky, gives good results in simple magnetic field
configurations, when the wave propagation directions are easily
predictable. For more complex magnetic field configurations, the
calculation of the characteristic directions is not an easy task.
The other alternative, \ie, sponge layer, consists in locating an
absorbing layer at the boundary to dissipate the wave energy and
prevent it from coming back to the physical domain. This strategy
is implemented in the code SLiM \citep{Cameron+etal2007}.
Absorbing layers give goods results only when the absorption is
gradual and needs a large amount of grid points. Thus, numerically
they are very costly. In our code we used yet another alternative,
the Perfectly Matched Layer \citep{Berenger1994}.

The Perfect Matched Layer (PML) is designed to absorb waves
without reflections. This method was first introduced by
\citet{Berenger1994} to absorb electromagnetic waves in numerical
solutions of Maxwell equations. Later it has been applied to Euler
equations \citep{Hu1996} and to simulations of acoustic waves in a
strongly stratified solar convection zone
\citep{Parchevsky+Kosovichev2007}. In our code we extend the
method to the full set of the MHD equations (Eqs.
\ref{eq:den1}--\ref{eq:ind1}). The MHD equations can be written
schematically, in conservative form, as:

\begin{equation}
\label{eq:pml} \frac{\partial {\bf u}}{\partial t}+\frac{\partial
{\bf F}({\bf u})}{\partial x} +\frac{\partial {\bf G}({\bf
u})}{\partial y}+\frac{\partial {\bf K}({\bf u})}{\partial z}={\bf
H}({\bf u}),
\end{equation}

\noindent where ${\bf u}\equiv[\rho_1, (\rho_0+\rho_1){\bf v_1},
e_1, {\bf B_1}]$ is the vector that contains the conserved
variables; the vectors ${\bf F}$, ${\bf G}$ and ${\bf K}$ are the
fluxes, whose expressions can be found in the system
(\ref{eq:den1}--\ref{eq:ind1}); and ${\bf H}$ represents the
source terms at right-hand side of the same equations. Inside the
PML, variables ${\bf u}$ are split into three components in such a
way that ${\bf u}={\bf u_1}+{\bf u_2}+{\bf u_3}$ and also ${\bf
H(u)}={\bf H_1(u)}+{\bf H_2(u)}+{\bf H_3(u)}$ and the system of
MHD equations is split into a set of three coupled, one
dimensional equations:

\begin{equation}
\label{eq:pml1} \frac{\partial {\bf u_1}}{\partial
t}+\frac{\partial {\bf F}({\bf u})}{\partial x}+\sigma_x(x){\bf
u_1}={\bf H_1}({\bf u}),
\end{equation}

\begin{equation}
\label{eq:pml2} \frac{\partial {\bf u_2}}{\partial
t}+\frac{\partial {\bf G}({\bf u})}{\partial y}+\sigma_y(y){\bf
u_2}={\bf H_2}({\bf u}),
\end{equation}

\begin{equation}
\label{eq:pml3}
\frac{\partial {\bf u_3}}{\partial t}+\frac{\partial {\bf K}({\bf u})}{\partial
z}+\sigma_z(z){\bf u_3}={\bf H_3}({\bf u}).
\end{equation}

These split Eqs. (\ref{eq:pml1}--\ref{eq:pml3}) are solved
independently in the PML, in contrast to the unsplit forms (Eq.
\ref{eq:pml}) which are solved in the physical domain.

An absorption term has been added to each equation. The
coefficients ${\sigma_x}(x)/{\sigma_y}(y)/{\sigma_z}(z)$ only
depend on x/y/z coordinate and are non-zero in the x/y/z PML
faces, respectively. Theoretically, a PML with constant absorption
coefficient produces no reflections for plane waves incident on a
flat interface for any angle of incidence and any frequency.
However, due to the finite difference implementation of the PML
equations in numerical calculations, reflections may appear when
$\sigma$ has a steep gradient \citep{Berenger1996}. To solve this
problem it is necessary to include smooth variations in the
absorption coefficients from small values at the interface between
the PML medium and the physical domain to large values at the
outer boundary. Following \citet{Hu2001}, good results are
obtained with absorption coefficients of the form:

\begin{equation}
\label{eq:sigmax} \sigma_x=\frac{a}{\Delta x}\Big (\frac{x-x_{\rm
PML}}{x_{\rm PML}}\Big )^2
\end{equation}

\begin{equation}
\label{eq:sigmay} \sigma_y=\frac{b}{\Delta y}\Big (\frac{y-y_{\rm
PML}}{y_{\rm PML}}\Big )^2
\end{equation}

\begin{equation}
\label{eq:sigmaz} \sigma_z=\frac{c}{\Delta z}\Big (\frac{z-z_{\rm
PML}}{z_{\rm PML}}\Big )^2
\end{equation}

\noindent where $\Delta x$, $\Delta y$ and $\Delta z$ are the
discretization steps, $a$, $b$ and $c$ are constants controlling
the damping amplitude, and $x_{\rm PML}$, $y_{\rm PML}$ and
$z_{\rm PML}$ are the thickness of the PML domain in each spatial
direction. In a typical calculation we need a PML with 10--15
grid points. The coefficients $a$, $b$ and $c$ depend on each
particular simulation and are proportional to the wave speed at
the boundaries. We locate PMLs at all boundaries of our simulation
domain. The results presented in \citet{Khomenko+etal2008a} show
that this strategy gives good results even for strong shocks.

The PML formulation, as presented above, can become unstable in
long simulation runs. According to the literature \citep[see,
\eg,][]{Hesthaven1998}, a high frequency noise filtering can
improve the long-time stability of the PML layer. We found that
this method works well for MHD waves. Applying a filter allows to
delay the effects of the possible instability the necessary time
to complete long enough simulation runs. Previously the PML layer
for MHD wave simulations was applied by
\citet{Parchevsky+Kosovichev2009}, though the stability of the PML
formulation for MHD equations was not discussed in this paper.

The code can also account for periodic and closed boundary
conditions.

\subsection{Parallelization}
%Paralelización

Parallelization has been done with MPI following a distributed
memory concept in which all data used by a processor are situated
on the memory partition accessible to it. Data are split in a
certain number of processors by means of a domain decomposition
scheme. The full numerical domain is divided into a set of three
dimensional subdomains, with communication between processors only
occurring at their common data boundaries. For this purpose, each
domain includes three layers of ``ghost'' cells at each boundary.
The 5-point stencil of the fourth-order scheme needs two cells
outside the subdomain, while for the filtering it is necessary to
include one more ``ghost'' layer. The boundaries of the subdomains
which are neighbors of other subdomains receive directly the
required information and store it in the ``ghost'' layers, while
when the boundary of the subdomain coincides with the global
boundary, the ``ghost'' layers are settled with the values imposed
by the boundary condition.

We have performed a number of standard numerical tests, as
well as tests specifically devised for wave propagation, to prove
the robustness of the numerical method and of the boundary conditions.
The results of these tests are given in Appendix \ref{sect:test}.

\section{3D propagation and transformation of waves in sunspot model}
\label{sect:analisis}

Below in this Section we discuss several simulations of the
propagation and transformation of MHD waves in a magnetostatic
sunspot model, excited by pulses with different periods and
locations. To facilitate the reading of this section, Table 1
summarizes the simulation runs. It gives the number of the
sub-section where the results are presented, the properties of the
driver, its location, horizontal source size, the duration of the
simulations and the numbers of the corresponding figures. In all
cases, to identify the different wave modes in three spatial
dimensions we use projections of the velocity into three
characteristic directions. To quantify the mode transformation we
calculate the acoustic and magnetic energy fluxes (see Section
\ref{sect:proyection}).

%%%%%%%%%%%%%%%%%%%%%%%%%%%%%%%%%%%%%%%%%%%%%%%òàáëèöà 1
\begin{table*}[t]
\begin{center}
\caption[]{\label{tab:simulations}
          { Summary of the simulation runs}}
\begin{tabular*}{0.6\textwidth}{@{\extracolsep{\fill}}cccccc}

\hline Section  & Driving & Location & $R_{\rm src}$  & Duration &
Figures
\\
 \hline
 4.4  &  50 s, harmonic &  $x=0$ Mm  & 150 km & 904 s  & 7--12 \\
 4.5  &  50 s, harmonic &  $x=-3$ Mm & 150 km & 1023 s & 13--17 \\
 4.6  &  180 s, harmonic & $x=0$ Mm  & 540 km & 930 s  & 18 \\
 4.7  &  300 s, wavelet & $x=0$  Mm  & 900 km & 1511 s & 19--21 \\
 4.7  &  300 s, harmonic & $x=0$ Mm  & 900 km & 1212 s & $-$ \\ \hline
\end{tabular*}
\end{center}
\end{table*}
%%%%%%%%%%%%%%%%%%%%%%%%%%%%%%%%%%%%%%%%%%%%%%%òàáëèöà 1

We use a vertical force in the momentum equation ${\bf S}(t)$ to
perturb a magnetostatic sunspot atmosphere in equilibrium and
study the waves generated by this perturbation. We have performed
several numerical simulations, all of them with the source
situated below the quiet photosphere at $z=-0.5$ Mm and $y=0$ Mm,
but with differences in the period, the horizontal $x$ location of
the source relative to the axis of the sunspot, as well as the
horizontal size of the source $R_{\rm src}$.

In the simulations described in Sect. \ref{sect:50s},
\ref{sect:50s_3000} and \ref{sect:180s} the temporal behavior of
the driver is harmonic is described by the expression:

\begin{equation}
\label{eq:source} S_z(r,t)= AP(r)\sin\frac{2\pi t}{\tau}
\end{equation}
In this equation, $A$ is the amplitude of the source,
$P(r)=\left[1- (r/R_{\rm src})^2\right]^2$ describes the source
shape, $R_{\rm src}$ is the source radius, $r=$ $ \sqrt{(x-x_{\rm
src})^2+(y-y_{\rm src})^2+(z-z_{\rm src})^2}$ is the distance from
the source center and $\tau$ is the period of the harmonic source.
$P(r)$ is zero if $r
> R_{\rm src}$. The x and y components of
$\textbf{S}$, $S_x(r,t)$ and $S_y(r,t)$, are set to zero.

In the simulations described in Section \ref{sect:300s} the
behavior of the driver is not harmonic in time, but rather has a
shape of Ricker wavelet:

\begin{equation}
S_z(r,t)= AP(r)\left(1-2\tau_0^2\right)e^{-\tau_0^2} \label{eq:wl}
\end{equation}

\noindent where $\tau_0 =\omega_0 t/2 -\pi$. Such driver produces
a spectrum of waves with a central frequency $\omega_0$
\citep{Parchevsky+Kosovichev2009}. We set $\omega_0 =3.33$ mHz, so
the spectrum of our driver resembles a solar one and covers a
broad range of frequencies.

This latter run is particularly interesting, as it allows us to
study the behavior of a realistic spectrum of solar waves in the
upper layers (photosphere and chromosphere) of a sunspot model,
including the propagation of individual wave modes and wave energy
fluxes. As far as we are aware of, no such investigation has been
performed as of today.

\subsection{Magnetostatic sunspot model} \label{sect:MHS}

Here we use a magnetostatic (MHS) model atmosphere in equilibrium
representative of a sunspot, adopted from
\citet{Khomenko+Collados2008}. This MHS model is a thick flux tube
with distributed currents, it is azimuthally symmetric and has no
twist. The variations of field strength and gas pressure are
continuous across the spot. At 40 Mm far from the sunspot axis the
model merges smoothly into a quiet Sun atmosphere taken from the
model S \citep{Christensen-Dalsgaard+etal1996} in the deep
sub-photosphere layers and continuing as a VAL-C model
\citep{Vernazza+Avrett+Loeser1981} in the photospheric and
chromospheric layers. The sunspot axis in the atmospheric layers
is given by the semi-empirical model of \citet{Avrett1981}. The
dimensions of our computational domain are 2.5 Mm in the vertical
direction and 15 Mm in each horizontal direction with a grid size
of $\Delta z=25$ km and $\Delta x=\Delta y=75$ km. The bottom
level is 1.25 Mm below $\tau_{5000}=1$ at the photospheric quiet
sun atmosphere, which was chosen as the zero level of the
coordinate $z$. The top level is 1.25 Mm above $z=0$. A PML layer
of 10 grid points was used at both bottom and top boundaries. With
this, the physical domain occupies from $z=-1$ Mm to $z=1$ Mm. The
axis of the sunspot is placed at the center of the simulation
domain. The magnetic field at the axis is about 900 G at $z=0$ Mm.

\subsection{Dispersion relations and propagation directions} \label{sect:dispersion}

In a three-dimensional situation like the one considered here,
three types of wave modes exist: fast and slow magneto-acoustic
waves and the Alfv\'en wave. Each of these modes is described by
its own dispersion relation, \ie\ relation between its temporal
frequency and the wave number $\omega=\omega ({\bf k})$. The modes
are characterized by their own phase velocity, ${\bf v}_{\rm
ph}=\omega /{\bf {k}}$, and the group velocity, ${\bf v}_{\rm
g}=\partial \omega /\partial {\bf k}$. The first one is parallel
to the direction of ${\bf {k}}$ and gives the wave front
propagation velocity. The second one defines the direction of the
energy propagation. The magnitudes and directions of ${\bf v}_{\rm
ph}$ and ${\bf v}_{\rm g}$ usually do not coincide.

In a 3D homogeneous unstratified atmosphere permeated by a
constant magnetic field ${\bf B_0}$, parallel to $z$ axis, and $y$
axis normal the plane defined by the direction of propagation
${\bf {k}}$ and ${\bf B_0}$, the generalized wave equation for the
perturbations decouples into two, defining the dispersion relation
for the three MHD modes:
\begin{equation}
\frac{\omega}{k}=v_A \cos\varphi , \label{eq:dispersion_alfven}
\end{equation}
\begin{equation}
\frac{\omega^2}{k^2}=\frac{1}{2}(v_A^2+c_S^2)\pm\frac{1}{2}\sqrt{(v_A^2+c_S^2)^2-
4v_A^2c_S^2\cos^2\varphi}, \label{eq:dispersion_fastslow}
\end{equation}

\noindent where $v_A=(B_0^2/\mu_0\rho_0)^{1/2}$ is the Alfv\'en
speed, $c_S=(\gamma p_0/\rho_0)^{1/2}$ is the sound speed, and
$\varphi$ is the angle between ${\bf k}$ and ${\bf B}_0$.

Equation (\ref{eq:dispersion_alfven}) describes an Alfv\'en wave, whose
associated perturbations in $\bf{v}_1$ and $\bf{B}_1$ are
transversal to $\bf{B}_0$ and ${\bf k}$ and the perturbations in pressure and
density are zero. For the propagation along the magnetic field the
phase velocity of this wave is equal to $v_A$. The energy of the
Alfv\'en mode propagates along magnetic field lines, according to
its group velocity.

The solutions with the plus and minus sign in Eq.
(\ref{eq:dispersion_fastslow}) are the fast and slow modes,
respectively. When one of the $c_S$ or $v_A$ is much higher than
the other, the dispersion equation (Eq. \ref{eq:dispersion_fastslow})
can be simplified to $\omega = k v_{\rm fast}$ for the fast mode
and $\omega = k v_{\rm slow} \cos\varphi =v_{\rm slow} {\bf k\cdot
B}_0/|B_0|$ for the slow mode, where $v_{\rm fast}$ is the highest
velocity between $v_A$ and $c_S$ and $v_{\rm slow}$ is the lowest.
The direction of the phase and group velocities of the fast mode
is ${\bf k}$ and their magnitude is either $c_S$ of $v_A$, so the
energy propagates in the same direction as the wave front. The
slow wave group velocity is directed along ${\bf B}_0$.
When $c_S$ and $v_A$ are similar, the behavior of the phase and
group velocities of the fast and slow modes deviates from this
simple picture, and can be retrieved from the derivation of ${\bf
v}_g$ from Eq. (\ref{eq:dispersion_fastslow}).

\begin{figure*}
\centering
\includegraphics[width=16cm]{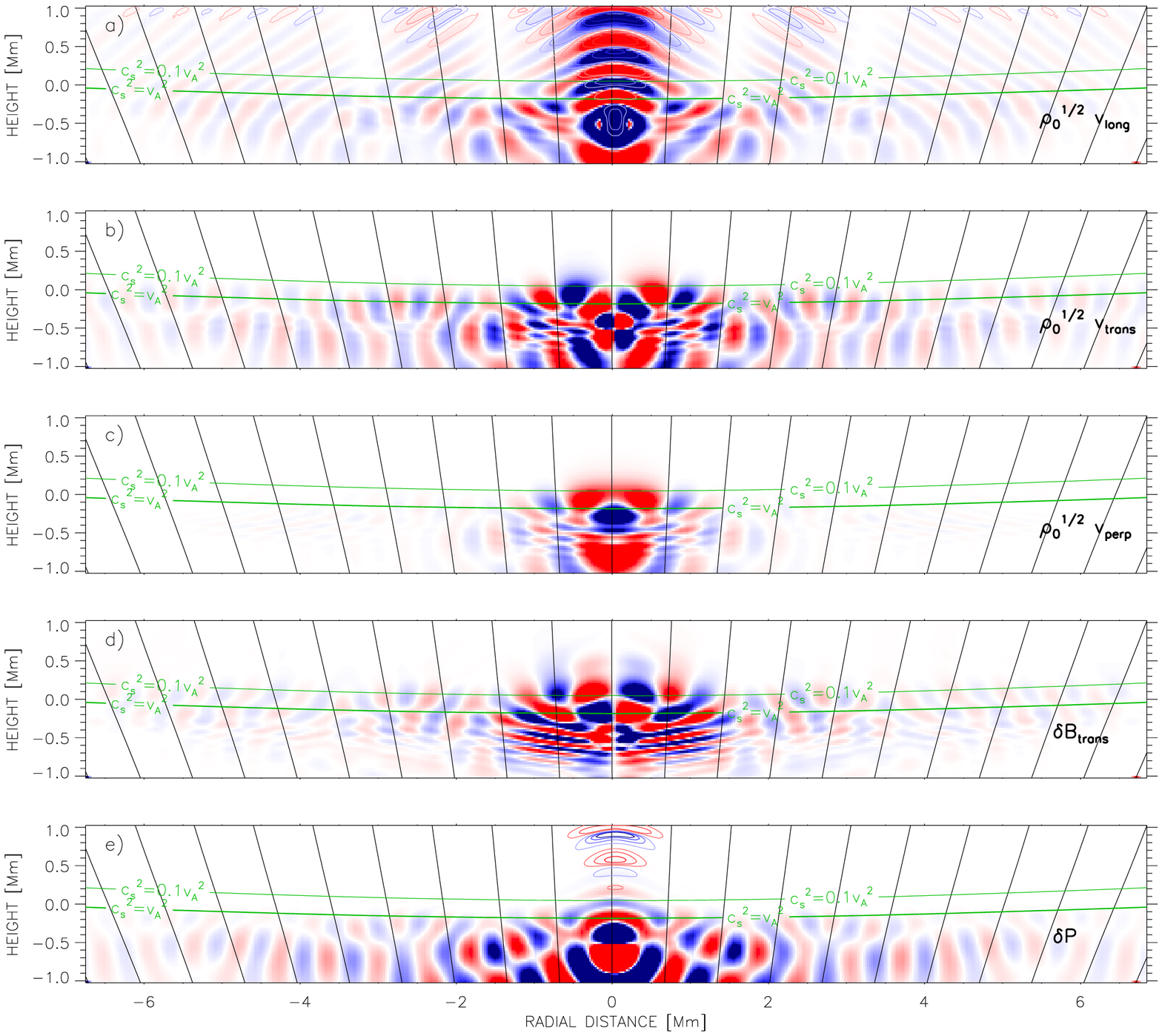}
\caption{Variations of the velocity in the direction $\hat{e}_{\rm
long}$ (a), $\hat{e}_{\rm trans}$ (b), and $\hat{e}_{\rm perp}$
(c), all of them scaled with factor $\sqrt{\rho_0}$ of the
unperturbed density; magnetic field in the direction $\hat{e}_{\rm
trans}$ (d), and pressure (e) at an elapsed time $t=820$ s after
the beginning of the simulations for the 50 s harmonic force
located at $x=0$ km, $y=0$ km and $z=-500$ km. All panels show the
plane $y=0$ km, except panel (c), which shows $y=400$ km. Black
inclined lines are magnetic field lines. Green lines are contours
of constant $v_A^2/c_S^2$. The image color coding is such that
blue colors represent lower values and red colors are higher
values with respect to the mean. Scaling in panels (a), (b), (c)
is the same. Figure (a) shows contours of equal longitudinal
velocity, and (d) shows contours of equal $p_1/p_0$.}
\label{fig:50arm}
\end{figure*}

\begin{figure}
\centering
\includegraphics[width=9cm]{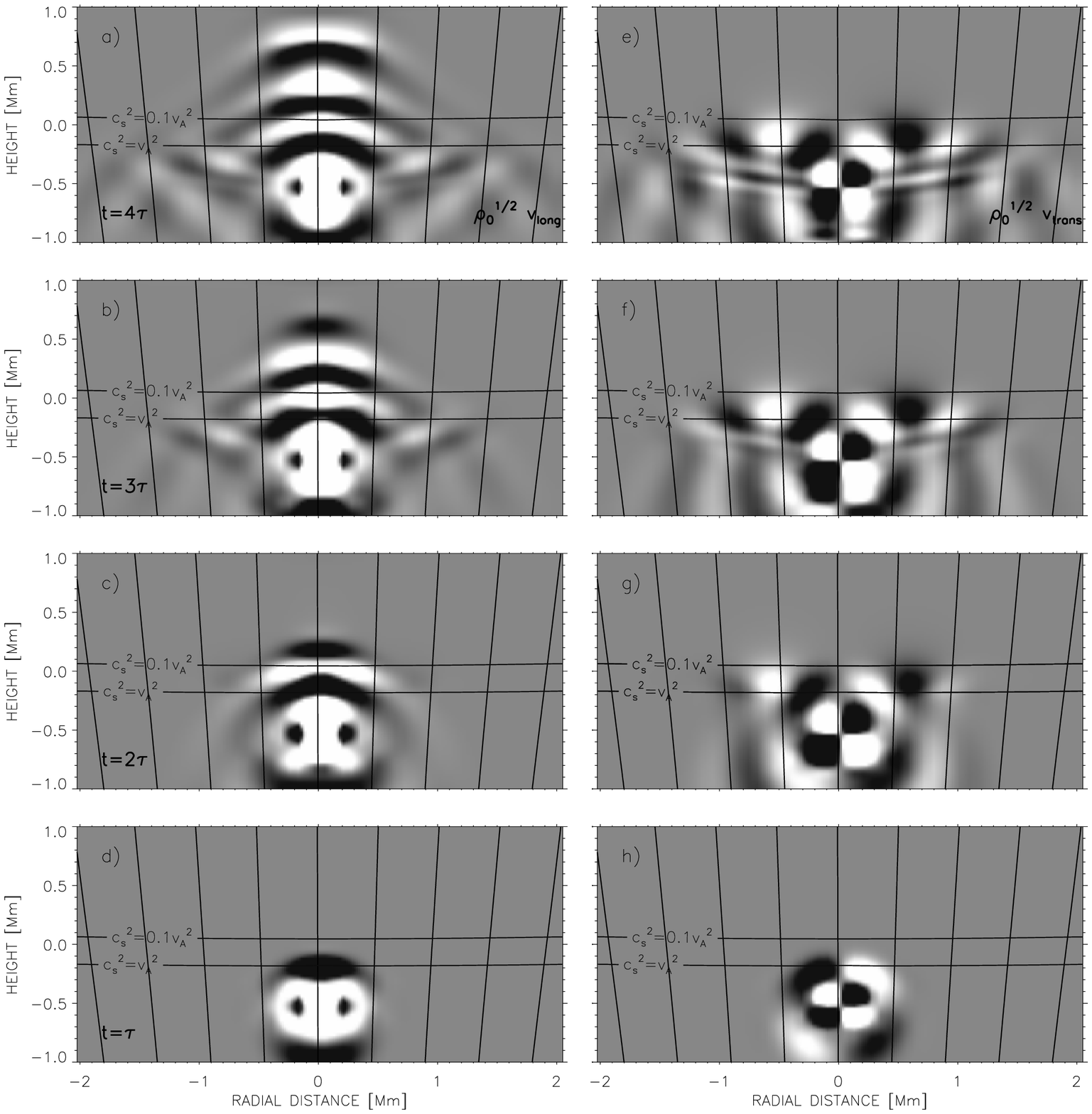}
\caption{Time evolution of $\sqrt{\rho_0}v_{\rm long}$ (left) and
$\sqrt{\rho_0}v_{\rm trans}$ (right) for the simulation with 50 s
harmonic driver. Time increases from bottom to top, with an
elapsed time between panels of a period ($\tau$). The image color
coding is such that black represents negative values and white
represents positive values. Horizontal lines are contours of
constant $v_A^2/c_S^2$.} \label{fig:50armt} %\vspace{-0.4cm}
\end{figure}

\subsection{Identification of the wave modes in simulations}
\label{sect:proyection}

In the case of a real atmosphere, the division into
pure wave modes is not so simple as in the idealized case described
above, as often no clear distinction between the modes can be done
neither physically nor mathematically. Even in the above simple
case the governing partial differential equation factors into a
single second order wave equation for the Alfv\'en mode and a
fourth order wave equation for the coupled fast-slow modes, so the
idea that there are three distinct modes may not always be
correct. However, the simplicity of this picture makes it
attractive and we will discuss the properties of the waves in
realistic atmospheres in terms of the three modes.

To help the identification of the wave modes in simulations, we
use the mode properties described above. We project the vectorial
quantities (velocity and magnetic field perturbations) into the
directions aligned/normal to the equilibrium magnetic field ${\bf
B}_0$.

At each location of the computational domain, we calculated the
projections of the $\bf{v_1}$ and $\bf{B_1}$ into the following
Cartesian directions:

\begin{equation}
\hat{e}_{\rm long}=[\cos\phi \sin\theta, \, \sin\phi \sin\theta,
\, \cos\theta],
\end{equation}

\begin{eqnarray}
\hat{e}_{\rm perp}=[ & - & \cos\phi \sin^2\theta \sin\phi, \,
1-\sin^2\theta \sin^2\phi, \, \\ \nonumber & - & \cos\theta
\sin\theta \sin\phi],\label{eq:eperp}
\end{eqnarray}

\begin{equation}
\hat{e}_{\rm trans}=[-\cos\theta, \, 0, \, \cos\phi \sin\theta].
\end{equation}

\noindent where $\theta$ is the magnetic field inclination from
the vertical and $\phi$ is the field azimuth, measured from the
$x-z$ plane. The direction of $\hat{e}_{\rm long}$ is along the
magnetic field ${\bf B}_0$. The direction of $\hat{e}_{\rm perp}$
is normal to the field and was chosen following
\citet{Cally+Goossens2008} as the asymptotic polarization
direction of the Alfv\'en mode in the low-$\beta$ regime. The last
component $\hat{e}_{\rm trans}=\hat{e}_{\rm
long}\times\hat{e}_{\rm perp}$ is set in the direction normal to
the other two.

We expect that in a region where $c_S > v_A$, the slow
magneto-acoustic mode will be identified in $\hat{e}_{\rm trans}$
projection of the velocity vector, while the fast magneto-acoustic
mode will be equally visible in all velocity components as is
propagates isotropically.
In a region where $c_S < v_A$, the slow magneto-acoustic mode will
be identified projected into $\hat{e}_{\rm long}$ direction, the
Alfv\'en mode projected into the $\hat{e}_{\rm perp}$ direction,
and the fast magneto-acoustic mode in the the direction normal to
these two.

To quantify the amount of energy contained in different wave modes
and to develop a measure of the mode transformation, suitable in
the case of complex magnetic field configurations like the one
considered here, we found it useful to calculate the wave energy
fluxes \citep{Bray+Loughhead1974}. The acoustic energy flux is
given by the expression:
\begin{equation}
{\bf F_{ac}}=p_1{\bf v}_1,
\label{eq:Fac}
\end{equation}
and magnetic energy flux is given by:
\begin{equation}
{\bf F_{mag}}={\bf B_1}\times({\bf v}_1\times {\bf B_0})/\mu_0.
\label{eq:Fmag}
\end{equation}

The acoustic energy flux contains the energy of the wave with
acoustic nature, which corresponds to the fast mode in the region
where $v_A < c_S$, and to the slow mode in the region where $v_A >
c_S$. In this region, the magnetic flux includes the fast and the Alfv\'en modes.
Since, as we will see in the next section, in the region above the
layer where $v_A = c_S$ the fast mode is refracted down towards
the photosphere, the magnetic energy which propagates upwards
along field lines must correspond to the Alfv\'en wave, making
possible the identification of this mode.

\subsection{Case of 50 s harmonic force located at the axis}
\label{sect:50s}

Figure \ref{fig:50arm} presents a two-dimensional snapshot of some
variables and Figure \ref{fig:50armt} gives the temporal evolution
of the projected velocities in the simulation run with the
harmonic 50 sec force located at the sunspot axis. The panels
(a--c) in Figs. \ref{fig:50arm} show the longitudinal, transversal
and perpendicular velocities scaled with a factor $\sqrt{\rho_0}$.
These magnitudes provide the square root of the kinetic energy
associated to the waves. Due to the strong density fall off, some
velocity perturbations at high layers have so low energy that
makes them indistinguishable in this representation. To make them
visible, we have plotted additional contours of constant velocity.
In the case of the pressure (Fig. \ref{fig:50arm}e), its drop with
height makes the absolute valued of the perturbations at the lower
layers much higher than at the upper layers. In this case, the
contours represent the ratio of constant $p_1/p_0$.

\subsubsection{Propagation below the surface}

The vertical force acts in a region where $c_S^2/v_A^2 \approx
9.1$ and it generates mainly an acoustic fast mode, whose
oscillations can be seen in the longitudinal velocity, pressure
and density snapshots in Fig.~\ref{fig:50arm}. This vertical
impulse produces initially a deficit in density and pressure at
the place where the source is located and, because of that,
horizontal motions also appear, creating a magnetic slow mode seen
in the transversal velocity and magnetic field snapshots in
Fig.~\ref{fig:50arm}.

At photospheric level, the longitudinal velocity has an amplitude
of about 200 \hbox{m s$^{-1}$}, the amplitude of transversal
velocity is 50 m s$^{-1}$ and the transversal magnetic field
oscillates with a maximum deviation from the equilibrium value of
4 G.

The temporal evolution given in Fig. \ref{fig:50armt} shows the fast
mode (acoustic in nature)  propagating in the deep layers upwards
to the region where $v_A \approx c_S$. It appears as a
perturbation in the longitudinal velocity (panels c--d).
Variations in longitudinal velocity are accompanied by acoustic variations of pressure and density (not shown in the
figure).
The slow mode (of magnetic nature) is visible in the transverse
% velocity variations (panels g--h). In this region $v_A < c_S$, so
the acoustic perturbation reaches the surface $v_A^2 = c_S^2$
earlier than the magnetic perturbation. In these deep layers, the
acoustic oscillations have a wavelength larger than the magnetic
ones.
As the fast (acoustic) wave propagates in the region $v_A < c_S$,
its energy is distributed in the three spatial dimensions and it
decreases away from the source as $1/r^2$. Once it reaches the
$v_A > c_S$ region, the energy redistribution in horizontal
directions is not so important because it is channeled along the
field lines and is only affected by the density falloff.

\begin{figure}
\centering
\includegraphics[width=9cm]{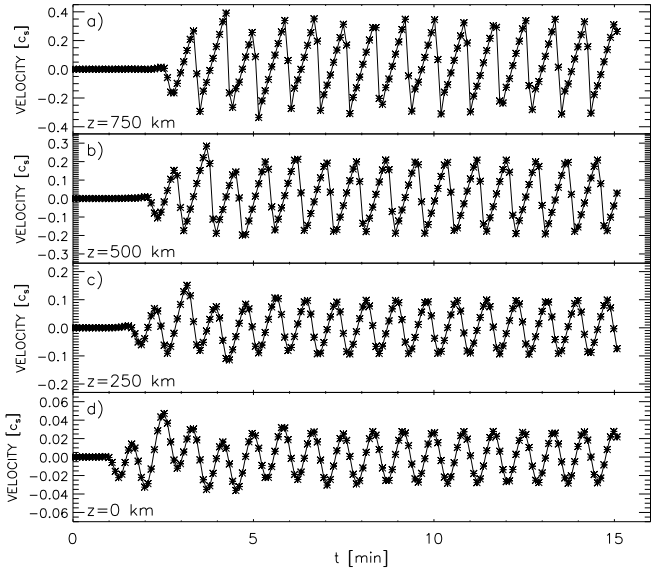}
\caption{Vertical velocity in the units of local sound speed at
the axis of the sunspot at several heights in the simulations with
50 s harmonic force. From bottom to top: $z=0$ km, $z=250$ km,
$z=500$ km and $z=750$ km.} \label{fig:shocks} %\vspace{-0.4cm}
\end{figure}

\subsubsection{Three-dimensional mode transformation}

When the waves reach the $v_A^2 = c_S^2$ layer from below, several
mode transformations take place in the simulation.

First of all, the fast acoustic mode moves from a region where
$v_A < c_S$ to another where $v_A > c_S$ keeping its acoustic
nature but changing from fast to slow mode. This transformation
can be seen in the snapshots of longitudinal velocity and relative
pressure in Fig. \ref{fig:50arm} (panels a and e) as the
wavefronts above the layer $v_A^2 = c_S^2$.
The atmosphere above $v_A^2 = c_S^2$ is dominated by the magnetic
field and this slow acoustic mode propagates upwards along field
lines (Fig. \ref{fig:50armt}, a--c). The amplitude of this wave
increases according to the density drop and it develops into
shocks above $z=500$ km. Figure \ref{fig:shocks} shows that the
oscillations in the vertical velocity develop a clear saw-tooth
shape with sudden decreases of the velocity followed by slower
increases. They present peak-to-peak variations of almost 8 km
s$^{-1}$ and their period is 50 s, the same period imposed by the
excitation pulse.

The second mode transformation is the acoustic fast mode which is
transmitted as a magnetic fast mode in the region $v_A > c_S$,
where the magnetic field dominates. The evolution of this magnetic
mode in the first 200 s of the simulation is clearly seen in Fig.
\ref{fig:50armt} (panels e--g) in the transversal velocity as the
wave which moves away from the axis of the sunspot just above the
surface $v_A^2 = c_S^2$. This mode is also visible in the
transversal magnetic field variations (Fig. \ref{fig:50arm}d).
Due to the rapid increase of the $v_A$ with height the fast
magnetic mode refracts and reflects back to the sub-photosphere,
showing a behavior similar to the two-dimensional case considered
by \citet{Khomenko+Collados2006}.

\begin{figure}
\centering
\includegraphics[width=9cm]{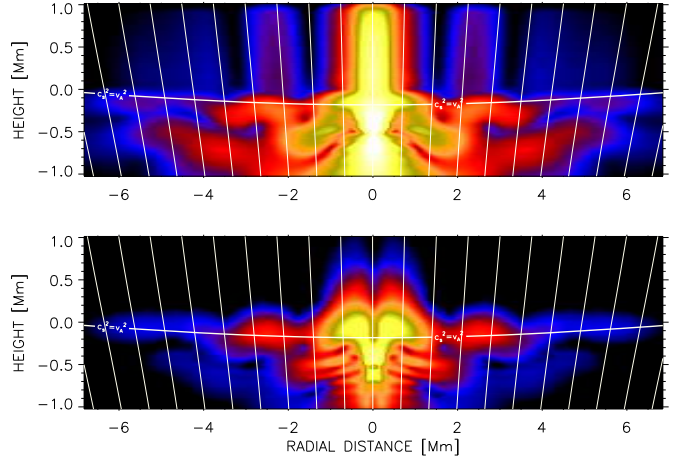}
\caption{Acoustic $(top)$ and magnetic $(bottom)$ flux for the
simulation with a 50 s harmonic driver located at the axis of the
sunspot averaged over the stationary stage of the simulations.
Horizontal white line is the height where sound speed and Alfv\'en
speed are equal. Vertical lines are magnetic field lines. The axis
are not to scale.} \label{fig:50arm_flux} %\vspace{-0.4cm}
\end{figure}

When the reflected magnetic fast wave reaches again the $v_A^2 =
c_S^2$ layer, it suffers two new transformations: a fast to slow
transformation, resulting in a magnetic wave in the $v_A < c_S$
region; and a fast to fast transmission, which produces a new
acoustic wave below $v_A = c_S$. First, we discuss the slow
magnetic wave. In Fig. \ref{fig:50arm} (panels b--d) it is clearly
visible in the transversal and perpendicular velocities and the
magnetic field variations, at horizontal locations inside a radius
of 1.5 Mm around the axis and at heights between $z=-1$ Mm and
$z=0$ Mm. Observing the temporal evolution at the beginning of the
simulation one can see that in Fig. \ref{fig:50armt}g the fast
magnetic wave above $v_A^2 = c_S^2$ has been refracted down and it
is located between the two horizontal lines which indicate
surfaces of constant $v_A^2/c_S^2$, but it has not arrived to the
lower one, so the new transformation has not been produced. In the
next time step (Fig. \ref{fig:50armt}f) the wave has already been
transformed in a slow magnetic mode in the region below $v_A^2 =
c_S^2$. The wavelength of this slow mode decreases as the wave
propagates to deeper layers because of the drop of the Alfv\'en
speed, which falls from 15 km s$^{-1}$ at $z=0$ Mm to 2 km
s$^{-1}$ at $z=-1$ Mm. Due to the higher density at the deeper
layers the amplitude of this wave also decreases as it propagates
down.

On the other hand, after the downward fast to fast transmission
has occurred from the refracted fast wave, another fast acoustic
wave appears in the region $v_A < c_S$. It is visible in
longitudinal velocity Fig. \ref{fig:50arm}a and in pressure in
Fig. \ref{fig:50arm}d below the layer $v_A^2 = c_S^2$. The
presence of this new acoustic mode can be checked comparing Fig.
\ref{fig:50armt}c and Fig. \ref{fig:50armt}b. In the latter one,
there is a new wave situated at both sides of the axis of the
sunspot at a radial distance between 0.5 and 1.5 Mm which can be
seen in longitudinal velocity. This mode appears after the
reflected fast mode in the region $v_A
> c_S$ reaches the surface $v_A^2 = c_S^2$. It propagates faster
than the slow magnetic wave mentioned before, with a speed close
to the sound speed, and its wavelength is larger than that of the
slow mode. It keeps the direction of the incidence of the fast
magnetic wave in the layer where $v_A^2 = c_S^2$, so it propagates
down with some inclination with respect to the vertical, moving
away from the axis.

In order to investigate the presence of the Alfv\'en mode in this
simulation (either before of after the transformations) we have
plotted in Fig. \ref{fig:50arm}c the velocity component in the
direction $\hat{e}_{\rm perp}$. As expected, this velocity
component has a node in the plane $y=0$ Mm. Thus, we present in
Fig. \ref{fig:50arm}c a vertical cut out of this plane, at 0.4 Mm
from the center of the computational domain. In this projection,
the Alfv\'en wave, if present, should appear as velocity
oscillation above $v_A^2 = c_S^2$ layer. The inspection of Fig.
\ref{fig:50arm}c shows that there are no oscillations  in the
magnetically dominated layers that can be identified an Alfv\'en
mode. Our conclusion is that when the wave driving occurs at the
axis of the sunspot, no conversion to Alfv\'en waves happens.

\subsubsection{Propagation in the upper atmosphere}

Velocity contours in Fig. \ref{fig:50arm}a show the presence of
some waves at a radial distance between $\pm 1.5$ and $\pm 6$ Mm
near the top of the computational domain. These waves are
specially tricky. From the first look at the figures and from the
time evolution of the snapshots one may have an impression (from
the inclination of the wave front) that they propagate across
field lines, opposite to the normal behavior of a slow acoustic
wave. However, we verified that their propagation speed is equal
to $c_S$, indicating their acoustic nature. The analysis of their
wave numbers indicates that the direction of propagation of these
waves is close to the inclination of the magnetic field lines.
Thus, we came to the conclusion that these waves are slow acoustic
waves created from the continuous transformation of the fast
acoustic mode which moves away from the driver across the field
lines below $v_A^2 = c_S^2$ (visible in Fig. \ref{fig:50arm}b in
transversal velocity and in Fig. \ref{fig:50arm}d in pressure
variations from a radial distance of $\pm 2$ Mm to $\pm 6$ Mm)
when it reaches this layer. As the wave front closer to the axis
of the sunspot gets to the surface $v_A^2 = c_S^2$ earlier, the
slow acoustic wave that it produces has the wave front inclined in
the direction to the axis. Due to this fact, it looks like its
propagation is across the field lines. This hypothesis also
explains the nodes located at a radial distance of $\pm 1.5$ and
$\pm 3$ Mm.

\begin{figure}
\centering
\includegraphics[width=9cm]{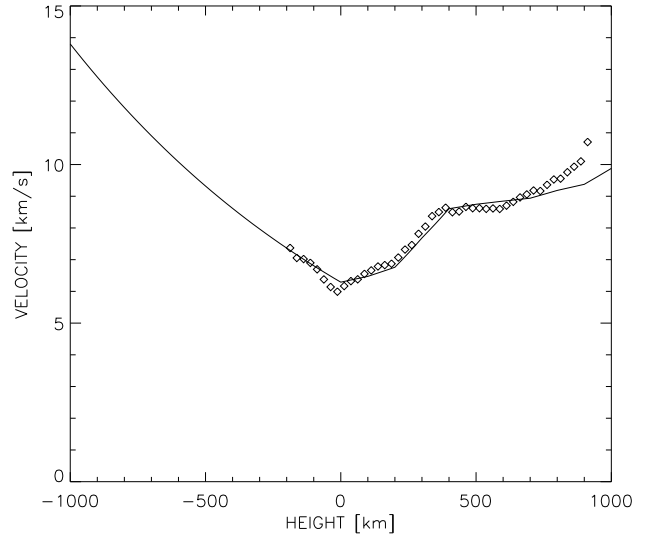}
\caption{Sound speed profile (solid line) at the axis of the
sunspot. Diamonds represent the phase velocity of the slow
acoustic mode measured from the simulations with 50 s harmonic
force.} \label{fig:cs} %\vspace{-0.4cm}
\end{figure}

\subsubsection{Acoustic and magnetic wave energy fluxes}

Figure \ref{fig:50arm_flux} shows the acoustic and magnetic fluxes
averaged over the stationary stage of the simulations. Around 95\%
of the flux at the location of the driver is acoustic flux and it
corresponds to the fast mode. At the center of the sunspot, where
the magnetic field is almost vertical, the fast to slow
transformation is very effective and the region above the layer
$v_A^2=c_S^2$ at the axis is dominated by the slow acoustic mode.
When the angle between the direction of propagation of the wave
and the magnetic field is different from zero, the fast to fast
transmission is produced and it forms the two lobes which are
visible in the magnetic flux just above the layer $v_A^2=c_S^2$.
The magnetic flux reaches a height of 0.5 Mm before the fast waves
are refracted back toward the photosphere. In the low-$\beta$
region, the fast magnetic mode has an important contribution from
$z=0$ Mm to $z=0.5$ Mm for radial distances below 2 Mm, except at
the axis of the sunspot. The transformation of the refracted fast
magnetic wave when it comes back towards the photosphere generates
acoustic as well as magnetic flux in the high-$\beta$ region,
corresponding to the new fast and slow modes, respectively, which
propagate downwards. As expected, we do not find any propagating
magnetic flux along the field lines, that may be associated with
an Alfv\'en mode.

\subsubsection{Slow acoustic mode phase velocities}

The phase velocity of a linear high-frequency slow mode wave in
a magnetically dominated region is equal to $c_S$. In our
simulation, a slow acoustic wave appears above the $v_A^2 = c_S^2$
layer produced after the mode transformation. This wave allows us
to check whether or not the velocity of waves involved in the
simulations in this complex sunspot model corresponds to that
expected from theoretical considerations.

Figure \ref{fig:cs} presents the results of this test. The solid
line in Figure \ref{fig:cs} shows the stratification of the sound
speed with height at the sunspot axis and diamonds indicate the
phase velocity of the slow mode wave measured at each grid point
from the simulations. Note that diamonds are only plotted at
heights above $z=-200$ km, after the mode transformation has been
completed. The velocity of the slow wave matches well the local
sound speed at heights from $z=-200$ km to $z=700$ km. Higher than
$z=700$ km the wave starts to propagate faster than $c_S$, since
the velocity amplitude of the wave approaches the sound speed and
non-linearities start playing an important role.

\begin{figure}
\centering
\includegraphics[width=9cm]{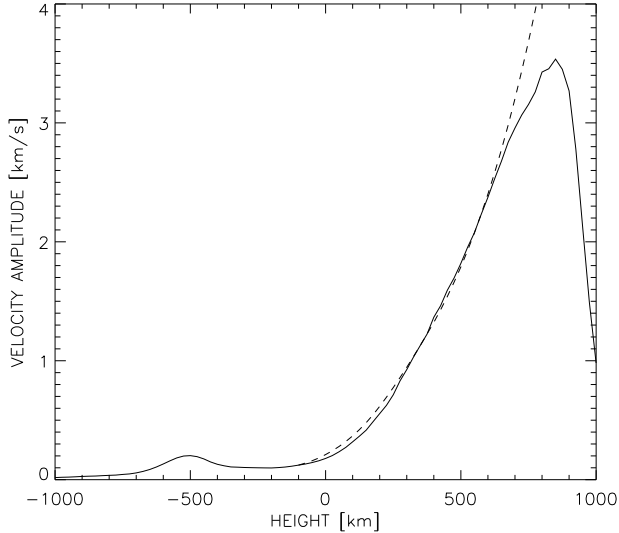}
\caption{Amplitude of the slow acoustic mode vertical velocity
(solid line) in the simulations with 50 s harmonic force. Dashed
lines gives the analytical curve for an acoustic-gravity wave with
50 s period \citep{Mihalas+Mihalas1984}.} \label{fig:amplitude}
%\vspace{-0.4cm}
\end{figure}

\subsubsection{Slow acoustic mode amplitudes}

As the slow acoustic wave propagates up in the atmosphere of the
sunspot the amplitude of the velocity oscillations increases due
to the density drop. The kinetic energy of this wave is
proportional to $\rho v^2$ and must be conserved. Thus, a decrease
of the density must be accompanied with an increase of the
velocity. The atmosphere of the sunspot is very complex and it
includes vertical and horizontal gradients in all the magnitudes.
Because of that there is no analytical expression for the
variation of the amplitudes of the waves with height. However, we
can compare the particular case of the wavefront of the slow
acoustic mode wave which propagates along the axis of the sunspot
with a case of a linear acoustic wave which propagates upwards in
a gravitationally stratified atmosphere permeated by a magnetic
field parallel to the direction of gravity. In this theoretical
case, the amplitude of the wave is given by
\begin{center}
\begin{equation}
\label{eq:amplitude} A(z)=A_0
\exp\left({\int_{z_0}^{z}{\frac{dz}{2H_0}}}\right),
\end{equation}
\end{center}

\begin{figure*}
\centering
\includegraphics[width=16cm]{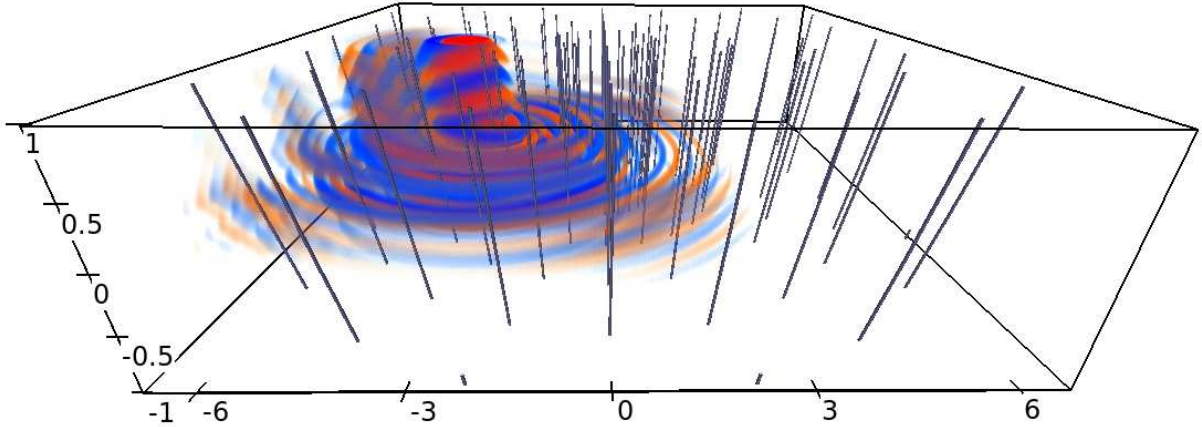}
\caption{Variations of $\sqrt{\rho}v_z$ at an elapsed time t=820 s
after the beginning of the simulations for the 50 s harmonic force
at 3 Mm from the axis of the sunspot. Grey inclined lines are
magnetic field lines. Blue colors represent upward movement while
orange/red colors are downward movement.}
\label{fig:50arm_3000_3D}
\end{figure*}

\begin{figure*}
\centering
\includegraphics[width=16cm]{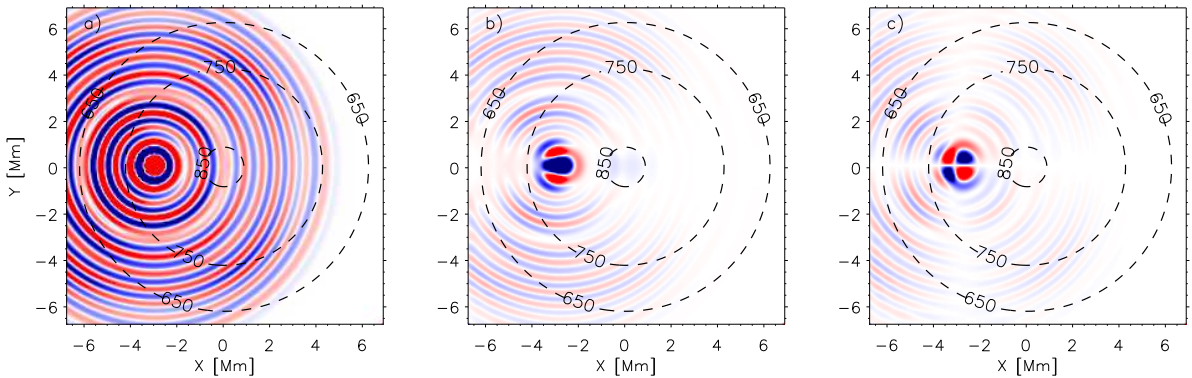}
\caption{Variations of the velocity in the direction $\hat{e}_{\rm
long}$ (a), $\hat{e}_{\rm trans}$ (b), and $\hat{e}_{\rm perp}$
(c), all of them scaled with factor $\sqrt{\rho_0}$ of the
unperturbed density, at $z=300$ km and at an elapsed time $t=820$
s after the beginning of the simulations for the 50 s harmonic
force located at $x=-3$ Mm , $y=0$ Mm and $z=-0.5$ Mm. The image
color coding is such that blue colors represent lower values and
red colors are higher values with respect to the mean. All images
have the same scale. Concentric dashed lines are contours of equal
magnetic field.} \label{fig:50arm_3000_v62}
\end{figure*}

In Fig. \ref{fig:amplitude} we show the amplitude of the vertical
velocity at each grid point at the axis of the sunspot (solid
line) for the stationary stage of the simulations. The force is
located at $z=-500$ km and the amplitude of the wave decreases
until it reaches $z=-200$ km. This initial decrease is due to the
part of the energy of the source that goes into other wave types
because of the mode transformations. More or less at this layer
the wave is transformed into a slow acoustic mode as it propagates
up while its amplitude increases. When the wave reaches the height
$z=850$ km its amplitude drops very fast as a consequence of the
large diffusivity that was imposed at high layers in order to
stabilize the numerical simulation. In this figure is also
overplotted the expected amplitude according to Eq.
(\ref{eq:amplitude}) (dashed line), starting from the height where
the wave has already been transformed into a slow acoustic mode in
the region \hbox{$v_A > c_S$}. From $z=-200$ km to $z=700$ km the
numerical amplitude agrees with the analytical one, while from
$z=700$ km to $z=850$ km the numerical amplitude is lower than the
analytical one. This happens because the wave develops into weak
shocks and the linear approximation for the amplitude increase is
no longer valid. Note that the amplitudes of the oscillations
(Fig. \ref{fig:amplitude}) as well as the phase velocity (Fig.
\ref{fig:cs}) show discrepancies with the linear theory at the
same heights.

\subsection{Case of 50 s harmonic force located off the axis}
\label{sect:50s_3000}

Fig. \ref{fig:50arm_3000_3D} gives a three-dimensional view of the
vertical velocity in the simulation run with 50 s harmonic force
located at $x=-3$ Mm off the sunspot axis. This figure clearly
shows the asymmetry of the wave front with respect to the axis. In
the lower part of the domain, the fast (acoustic) waves can be
appreciated propagating in circles away from the source with a
visibly lower amplitude toward the axis. In the upper part of the
domain, slow (acoustic) waves are the dominating ones, propagating
along the inclined magnetic field lines.

\begin{figure*}
\centering
\includegraphics[width=16cm]{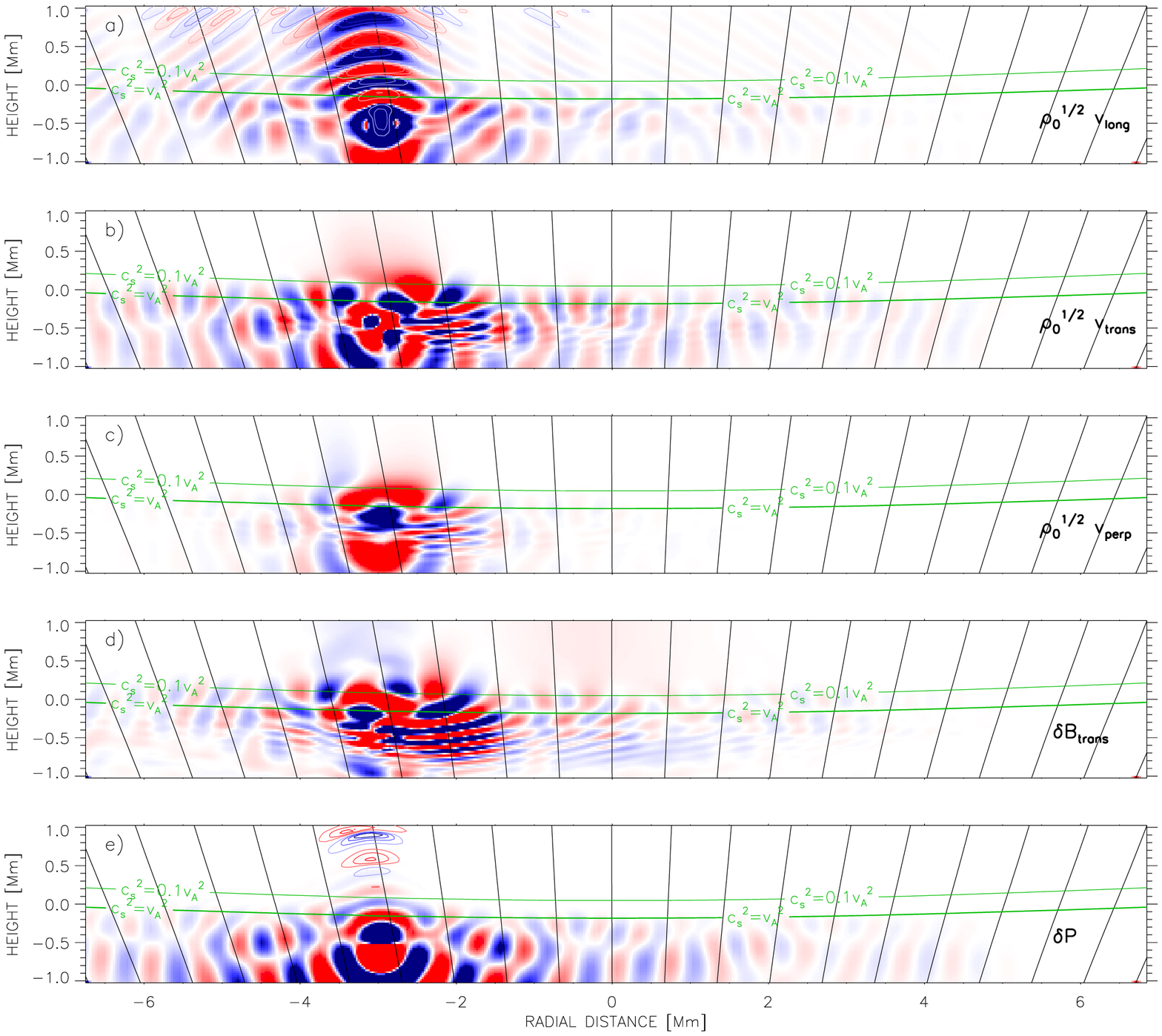}
\caption{Variations of the velocity in the direction $\hat{e}_{\rm
long}$ (a), $\hat{e}_{\rm trans}$ (b), and $\hat{e}_{\rm perp}$
(c), all of them scaled with factor $\sqrt{\rho_0}$ of the
unperturbed density; magnetic field in the direction $\hat{e}_{\rm
trans}$, (d) and pressure (e) at an elapsed time t=820 s after the
beginning of the simulations for the 50 s harmonic force located
at $x=-3$ Mm off the sunspot axis. The format of the figure is the
same as Fig. \ref{fig:50arm}.} \label{fig:50arm_3000}
%\vspace{-0.4cm}
\end{figure*}

Fig. \ref{fig:50arm_3000_v62} presents the projected velocities in
the three characteristic directions at the horizontal cut of the
simulation domain taken at the middle photosphere at $z=300$ km.
Fig.  \ref{fig:50arm_3000} shows the snapshots of some variables
in the vertical cut through the domain. Both figures correspond to
the same time moment of the simulations at $t=820$ s. These
simulations have many features in common with the previously
considered case of the driving force located at the sunspot axis.
A set of fast (acoustic) and slow (magnetic) modes is generated
below the layer $v_A^2 = c_S^2$, propagating upwards and suffering
several transformation after reaching this height. Similar to the
previous case, slow (acoustic) and fast (magnetic) modes are
produced after the mode transformation in the magnetically
dominated upper atmosphere. The conversion to slow and fast modes
in the low-$\beta$ region only presents slight changes in
comparison with the simulation with the driver placed at the axis.
One of these changes is the presence of an asymmetry with respect to
the axis. For example, the transformation of the downward
propagating refracted fast (magnetic) mode into the slow
(magnetic) mode below the surface (Fig. \ref{fig:50arm_3000},
panels b--d) presents such asymmetry. Due to the particular
combination of the field inclination and the direction of
propagation of the refracted fast mode, this transformation is
much more efficient on the right from the source in the direction
toward the axis (locations between $x=-3$ and $-1$ Mm at $z
\approx -0.5$ Mm).

The most important changes are present in the velocity component
in the direction $\hat{e}_{\rm perp}$
(Fig.~\ref{fig:50arm_3000_v62}c). At height of $z=300$ km, this
component shows variations that do not correspond to the fast
(magnetic) mode. The latter has been already reflected down. As
demonstrated by the analysis of the magnetic energy flux below,
the variations observed in the snapshot of $\hat{e}_{\rm perp}$ at
heights above $z \approx 300$ km at locations between $x=-4 \div
-2$ Mm and $y=-1 \div 1$ Mm correspond to the Alfv\'en wave.

\begin{figure}
\centering
\includegraphics[width=9cm]{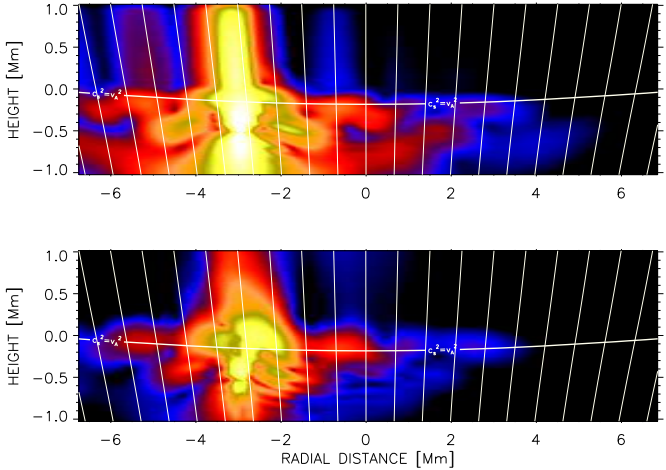}
\caption{Average acoustic $(top)$ and magnetic $(bottom)$ flux for
the 3D simulation with a 50 s harmonic driver located at 3 Mm from
the axis of the sunspot averaged over the stationary stage of the
simulations. The format is the same as Fig. \ref{fig:50arm_flux}.}
\label{fig:50arm_3000flux}
\end{figure}

The acoustic and magnetic fluxes are shown in Fig.
\ref{fig:50arm_3000flux}. These fluxes, in general, show a pattern
similar to the case of driving at the axis, but some important
asymmetry is also present. The acoustic flux of the slow mode in
the low-$\beta$ region is oriented in the direction of the field
lines and it is slightly lower than in the previous simulation due
to the larger angle between the direction of propagation of the
fast mode before reaching the layer $v_A^2=c_S^2$ and the magnetic
field ($\varphi$).
An important fraction of the magnetic flux appears in the lobe
located at $v_A^2=c_S^2$  line in the direction toward the axis of
the sunspot present due to more efficient fast to fast mode
transmission with increasing $\varphi$ \citep{Cally2005}. Compared
to Fig.~\ref{fig:50arm_flux} there is also more magnetic flux
present in the upper part of the atmosphere above $v_A^2=c_S^2$,
apparently directed along the magnetic field lines. It is not
clear, however, whether this flux corresponds  to fast, not yet
completely reflected wave, or to the Alfv\'en wave. To clarify
this issue we have calculated the magnetic flux following Eq.
(\ref{eq:Fmag}), but using only velocity and magnetic field
projections in the direction $\hat{e}_{\rm perp}$. We expect that
the longitudinal projection of this quantity gives us indications
about the presence of the propagating Alfv\'en waves.

\begin{figure}
\centering
\includegraphics[width=9cm]{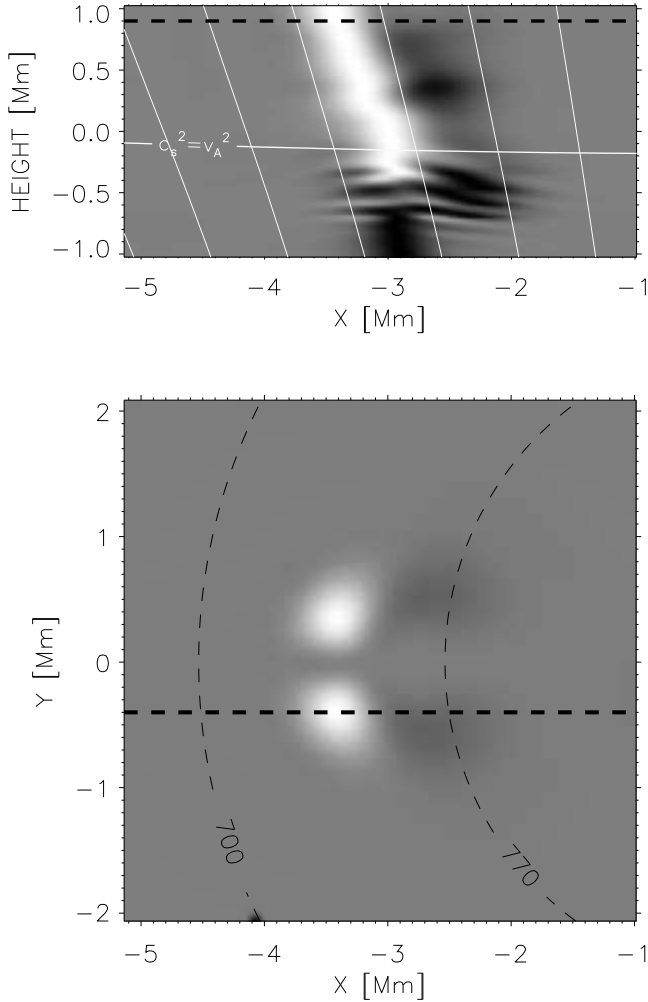}
\caption{Magnetic flux of the Alfv\'en mode. \emph{Top:} Vertical
cut in the plane $y=-0.4$ Mm, normalized at every height. Vertical
white lines are magnetic field lines and horizontal white line is
the layer where $c_S^2=v_A^2$. \emph{Bottom:} Horizontal cut in
the plane $z=0.9$ Mm. Thin dashed lines are contours of equal
magnetic field. In both panels, thick dashed lines mark the
location of the other plot.} \label{fig:flux_alfven}
\end{figure}

Figure \ref{fig:flux_alfven} illustrates the result. Top panel
corresponds to a vertical cut in the plane $y=-0.4$ Mm, normalized
at every height to its maximum value at this height. The bottom
panel is a horizontal cut in the plane $z=0.9$ Mm. The white
colors (positive flux) mean upward energy propagation, while the
black colors (negative flux) mean downward energy propagation.
Indeed these plots reveal the energy flux associated to the
Alfv\'en wave, which clearly propagates upwards along the field
lines. The Alfv\'en mode has a node at the plane $y=0$ Mm, where
the driver was located, so conversion to this mode is only
produced when the wave vector forms a certain angle with the
magnetic field (different from zero). This result is in
qualitative agreement with the recent investigation of the
conversion to Alfv\'en waves by \citep{Cally+Goossens2008}.
However, even at the location where the contribution of the
Alfv\'en wave energy flux to the total energy flux is maximum, its
flux is still around 20 times lower than the acoustic flux at this
location.
So it means that in the sunspot magnetic field configuration and
for the driver location considered here, the transformation from
fast (acoustic) to Alfv\'en wave is much less effective than the
transformation to the slow (acoustic) wave in the magnetically
dominated upper atmosphere.

\subsection{Case of 180 s harmonic force located at the axis}
\label{sect:180s}

\begin{figure}
\centering
\includegraphics[width=9cm]{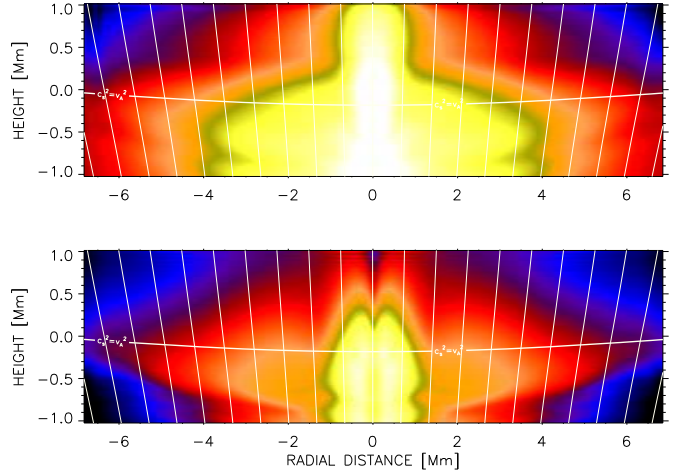}
\caption{Acoustic $(top)$ and magnetic $(bottom)$ flux for the
simulation with a 3 min harmonic driver located at the axis of the
sunspot averaged over the stationary stage of the simulations. The
format is the same as Fig. \ref{fig:50arm_flux}.}
\label{fig:180arm_flux}
\end{figure}

This simulation shows many features which also appeared in the 50
s harmonic case, since the response of the magnetic atmosphere is
quite similar at both frequencies because they are above the
cut-off frequency. So, in this case we omit the discussion of the
mode transformation and only show results of the acoustic and
magnetic flux calculations. These fluxes are presented in Fig.
\ref{fig:180arm_flux}.

This figure is very similar to Fig.~\ref{fig:50arm_flux}, except
for much larger wavelength of the fluctuations, in agreement with
the larger temporal period of waves. The average acoustic flux
shows that at the axis of the sunspot the fast to slow mode
transformation is also effective for waves with 180 s period.
After the transformation, slow (acoustic) waves propagate acoustic
energy upwards.
The fast mode visible in the magnetic flux above the $v_A^2=c_S^2$
also has longer wavelenghts than in the 50 s harmonic simulation.
The magnetic flux vanishes at the axis of the sunspot at high
layers. Due to the reflection of the fast (magnetic) wave, there
is no magnetic flux above the certain height, also away from the
axis. In this simulation we find no traces of the Alfv\'en mode.

\begin{figure*}[!ht]
\centering
\includegraphics[width=16cm]{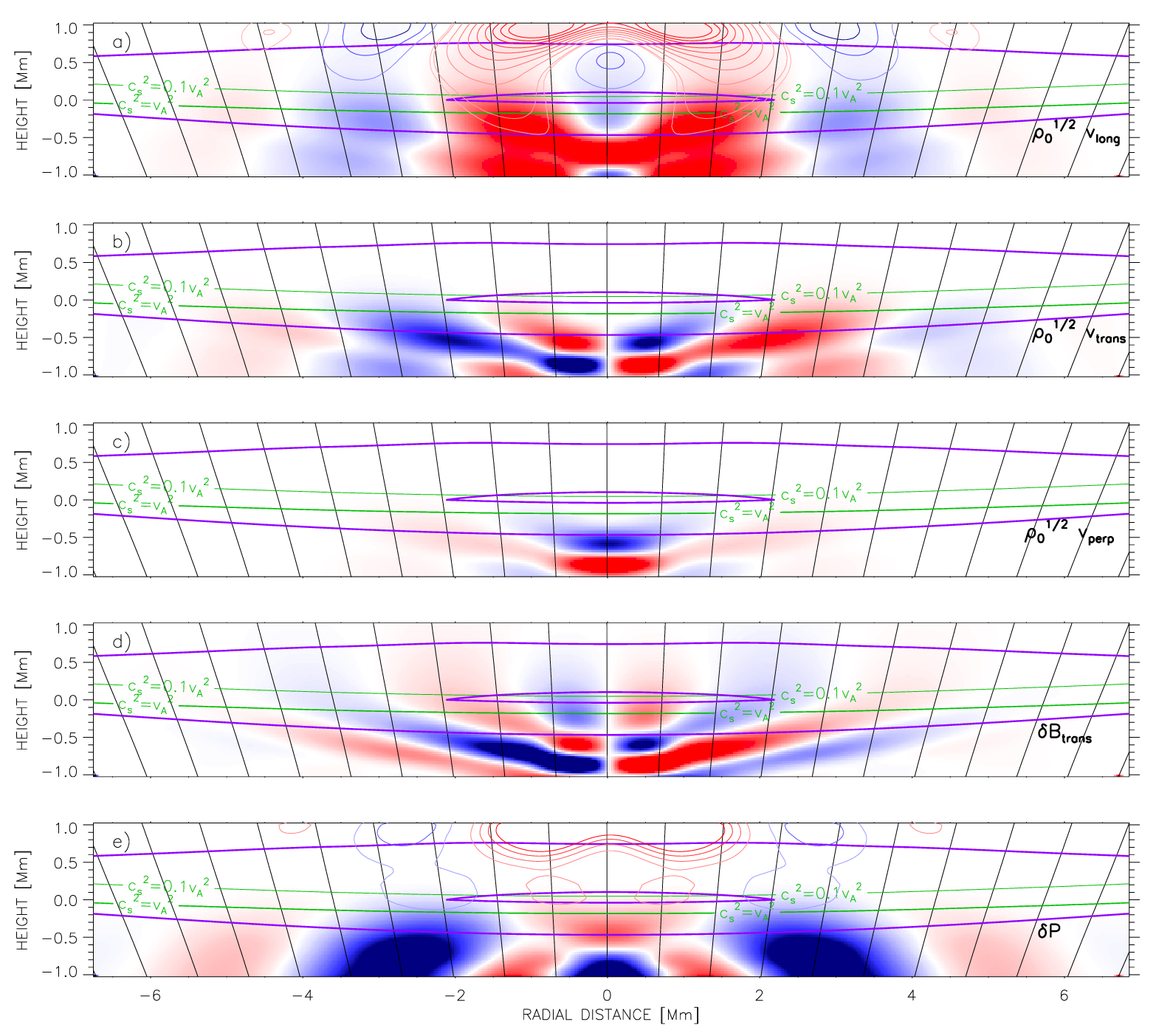}
\caption{Variations of the velocity in the direction $\hat{e}_{\rm
long}$ (a), $\hat{e}_{\rm trans}$ (b), and $\hat{e}_{\rm perp}$
(c), all of them scaled with factor $\sqrt{\rho_0}$ of the
unperturbed density; magnetic field in the direction $\hat{e}_{\rm
trans}$ (d), and pressure (e) at an elapsed time t=820 s after the
beginning of the simulations for the 300 s wavelet force located
at $x=0$ km, $y=0$ km and $z=-500$ km. The format of the figure is
the same as Fig. \ref{fig:50arm}. Violet lines represent contours
of equal cut-off frequency. The inner one is $\nu_c=5.6$ mHz and
the outer one is $\nu_c=4$ mHz.} \label{fig:300wav}
%\vspace{-0.4cm}
\end{figure*}

\subsection{Case of 300 s wavelet force located at the axis}
\label{sect:300s}

Figure \ref{fig:300wav} shows vertical snapshots of several
variables after 820 s of the simulation with the driver emitting a
spectrum of waves with a central frequency at 3.33 mHz given by
the Eq. (\ref{eq:wl}).

According to the stratification of the atmosphere, at the axis of
the sunspot at $z=-700$ km the cut-off frequency is $\nu_c=3.3$
mHz. It increases with height reaching a maximum at the height of
$z=0$ km, where its value is $\nu_c=5.8$ mHz. It means that 5
minute acoustic waves can not propagate upwards above $z=0$ Mm,
since they are evanescent in the vertical direction, and can only
propagate horizontally. Therefore, the cut-off frequency is a
critical value for the wave propagation in this case, and the
behavior of waves below and above the cut-off layer is expected to
be completely different.

The source located at $z=-500$ km drives waves with frequencies
below as well as above $\nu_c$, but most of its energy goes to the
band around 3.33 mHz. This generates a fast acoustic wave with an
amplitude below 100 \hbox{m s$^{-1}$} and 5 minute period, which can
propagate upwards only until the height $z=0$ Mm. Waves with this
frequency are evanescent at higher layers and their vertical
wavelength occupies the whole upper part of the simulated
atmosphere. The amplitude of their longitudinal velocity slightly
increases with height.

\subsubsection{Three-dimensional mode transformation}

Part of the energy of the driver which reaches the surface $v_A^2
= c_S^2$ is transformed into a fast wave above this height. Due to
its magnetic nature it becomes unaffected by the cut-off
frequency. As in the previous simulations, the transversal
velocity in Fig. \ref{fig:300wav}b and the transversal magnetic
field in Fig. \ref{fig:300wav}d show that the fast magnetic wave
in the region $v_A > c_S$ is reflected because of the gradients in
the Alfv\'en speed. Once the wave comes back to the
sub-photospheric layers below $v_A^2 = c_S^2$, it keeps its
magnetic nature and propagates downwards in a form of a slow wave
with decreasing wavelength due to the drop of the Alfv\'en speed,
but keeping its 5 minute period.

The maximum cut-off frequency at the axis in this sunspot model is
\hbox{$\nu_c=5.8$ mHz}, so waves with higher frequencies can still
propagate upwards through the atmosphere. The fast acoustic modes
generated by the driver with frequencies higher than $\nu_c$ are
transformed into propagating slow acoustic modes in the region
above $v_A^2 = c_S^2$. The contours in Fig. \ref{fig:300wav}a
represent constant longitudinal velocity. At a height around
$z=900$ km, the longitudinal velocity has maximum power at 3
minute period, which corresponds to the frequency above $\nu_c$
receiving more energy from the driver.

\begin{figure}
\centering
\includegraphics[width=9cm]{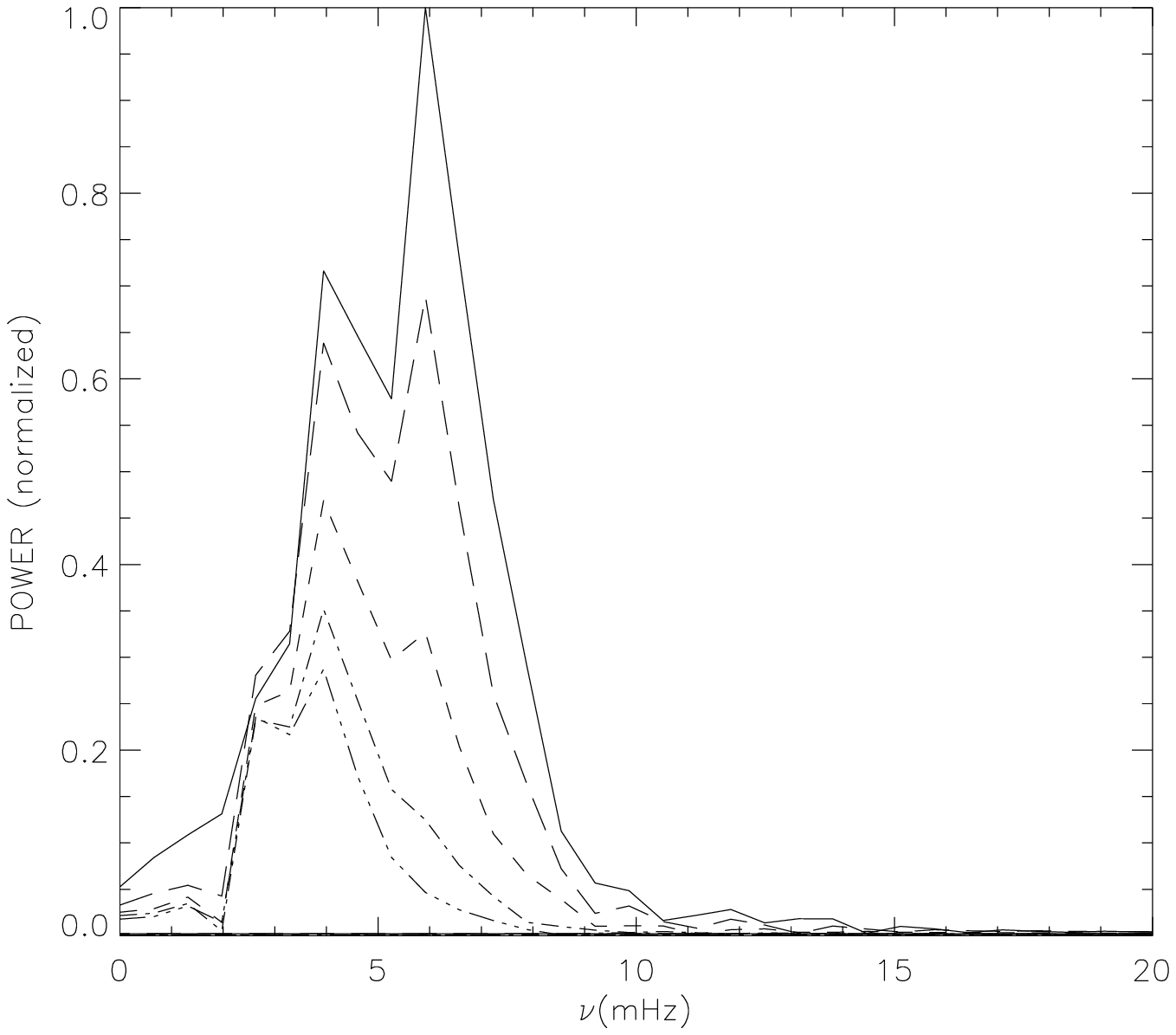}
\caption{Power spectra at different heights at the axis of the
sunspot for the simulations with 300 s wavelet force, normalized
to the maximum power at the highest height. From bottom to top:
$z=-250$ km, $z=0$ km, $z=250$ km, $z=500$ km and $z=750$ km.}
\label{fig:potencia_z} %\vspace{-0.7cm}
\end{figure}

\subsubsection{Frequency change with height}

The increase of the amplitude of 3-minute waves according to the
drop of the density (compared to much weaker increase of the
amplitude of the evanescent 5-minute waves)  leads to the power
spectrum at chromospheric heights dominated by 3-minute waves.
There, their amplitudes reach almost 400 \ms. They do not develop
into a saw-tooth waves because at the photosphere the driver
generates low power at this frequency band and their amplitude
increase is not enough to produce significant non-linearities.

Figure \ref{fig:potencia_z} shows the power spectra at different
heights, from $z=-250$ to $z=750$ km. At $z=750$ km it shows a
clear peak at $\nu =5.9$ mHz, corresponding to a period of 170 s.
At the lower layers, a peak below 4 mHz dominates. One can see in
this figure how the amplitude of the low-frequency peak increases
with height. However, this increase is weaker than the one of the
peak at high-frequency around  $\nu \approx 6$ mHz. Due to this
behaviour, the high layers are dominated by oscillations with
period around 3 minutes.

A simulation with a 300 s harmonic driver was also performed. In
this case high layers develop evanescent 5 minutes period waves,
but there is no trace of 3 minutes oscillations at the
chromosphere, so power in this frequency band can not be produced
if these frequencies are not excited by the driver. Thus, we can
conclude that the mechanism that produces the change in frequency
of oscillations in the umbra from the photosphere to the
chromosphere is the linear propagation of waves with 3 minute
power which come directly from the photosphere and dominate over
the evanescent long period waves. This conclusion goes in line
with the results of the observational study of sunspot waves
simultaneously at the photosphere and the chromosphere by
\citet{Centeno+etal2006a}.

\begin{figure}
\centering
\includegraphics[width=9cm]{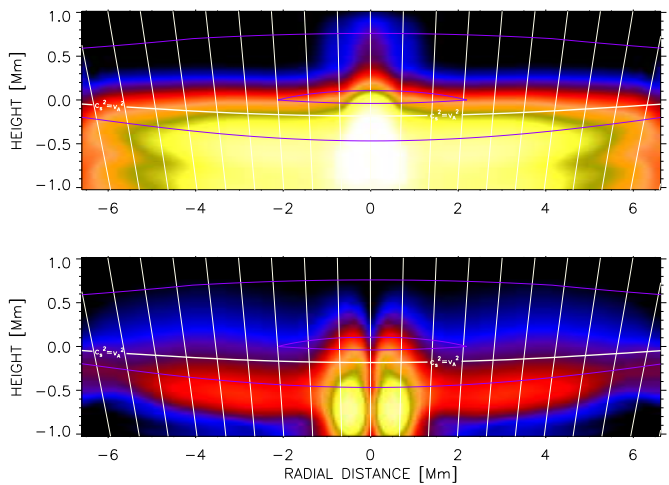}
\caption{Acoustic $(top)$ and magnetic $(bottom)$ flux for the
simulation with a 5 min wavelet driver located at the axis of the
sunspot averaged over the stationary stage of the simulations. The
format is the same as Fig. \ref{fig:50arm_flux}. Violet lines
represent contours of equal cut-off frequency. The inner one is
$\nu_c=5.6$ mHz and the outer one is $\nu_c=4$ mHz.}
\label{fig:300wav_flux} \vspace{-0.4cm}
\end{figure}

\subsubsection{Acoustic and magnetic wave energy fluxes}

The magnitudes $\sqrt{\rho_0}v_{\rm long}$, $\sqrt{\rho_0}v_{\rm
tran}$ and $\sqrt{\rho_0}v_{\rm perp}$ plotted in Fig.
\ref{fig:300wav} (panels a--c), respectively, show that most of
the kinetic energy remains in the photosphere and below. Most of
the energy introduced by the driver goes into the waves in the 5
minute band. Propagating to higher layers, they form evanescent
waves or are transformed into fast magnetic mode waves. The first
ones do not carry energy, while the second ones are reflected back
to the photosphere. Thus, waves in the 5 minute band can not
supply energy to the chromosphere, if the driving force is located
at the sunspot center.

This case differs from the simulations with shorter periods
described in Sects. \ref{sect:50s}, \ref{sect:50s_3000} and
\ref{sect:180s}, where the slow acoustic wave transports to the
high layers part of the energy injected by the driver (or, in the
case of 50 s off-axis driver, a smaller part of the energy is also
transported upwards in the form of the Alfv\'en wave).
In the simulation with the 5 minute wavelet force located at the
axis, only the waves with frequencies higher than $\nu_c$ can
provide energy to the chromosphere, and they represent a small
fraction of the energy introduced by the driver.

The acoustic flux in Fig. \ref{fig:300wav_flux} shows that most of
the energy keeps below the layer $c_S^2=v_A^2$. The wavelet mainly
drives 5 min power in a fast acoustic mode, and since it is
evanescent, it does not propagate energy upwards and this 5 min
power is distributed horizontally. Only waves with frequency
higher than the cut-off frequency are transformed into slow
acoustic modes in the low-$\beta$ region and carry energy to the
chromosphere. They correspond to the low acoustic flux which
appears at the center of the sunspot between $z=0$ Mm and $z=1$
Mm. The high-$\beta$ region contains the energy of the slow
magnetic modes generated from the secondary transformation of the
reflected fast magnetic modes. In the magnetic flux, the energy of
these slow modes appears in red for radial distances below 6 Mm.
Note that, interestingly, the maximum of this energy is not
located just at the $c_S^2=v_A^2$ line it was in the case of 50 s
simulations, but it located below this line. This new location
follows the line of constant $\nu_c=4$ mHz meaning that the
cut-off effects influence also the penetration of the magnetic
energy into the higher layers.
The fast mode waves in the high-$\beta$ region contribute to the
high acoustic flux there. Like in all the simulations with the
driver located at the center of the sunspot, magnetic flux along
field lines is negligible, and there is no conversion to Alfv\'en
waves.

\subsubsection{Propagation in the upper atmosphere}

Contours in Fig. \ref{fig:300wav}a in the region above $v_A^2 =
c_S^2$ (upper part of the domain) far from the axis of the sunspot
show longitudinal waves with 5 minute period which apparently move
across field lines at the sound speed. Similar to Sect.
\ref{sect:50s}, we tried to analyze their wave number and phase
velocity behavior with height. However, the wavelength of these
waves in the upper layers is comparable to the size of our
simulation domain in the vertical direction ($\sim$ 1 Mm above
\hbox{$v_A^2 = c_S^2$}). Because of that, we can not be completely
sure if these waves propagate along or across the field lines.

Another difficulty to understand the behavior of these waves lies
in the fact that their frequency is below the cut-off frequency
$\nu_c$ of the atmosphere. In principle, the cut-off frequency in
the magnetically dominated atmosphere is lowered for the acoustic
wave propagating along inclined field lines. We calculated the
cut-off frequency taking into account the effect of the field
inclination. We found this inclination to be insufficient to
reduce the cut-off frequency enough to allow for the propagation
of the 5-minute waves. Simulations in a larger spatial domain
(both in the horizontal and vertical direction) will be needed in
the future to clarify the nature of these waves.

\section{Conclusions}
\label{sect:conclusiones}

In this paper, we have addressed the problem of the
three-dimensional wave propagation and mode transformation of the
MHD waves in the upper atmosphere (photosphere and chromosphere)
of a sunspot model, by means of numerical simulations. We have
presented our code for the calculation of the response of a
magneto-static structure in equilibrium to an arbitrary
perturbation. This code is specially designed to model the wave
propagation in solar magnetic structures with non-trivial magnetic
field configurations. We performed several tests proving the
correct code performance.
The comparison of the numerical tests with the analytical
solutions (when they are available) or with the results from other
codes (in other cases) shows a precise agreement and demonstrates
the robustness of the numerical method applied  in this code
according to several aspects: the velocities of wave propagation,
the resolution of shocks and the conservation of energy. The
performance of the artificial diffusivity makes it possible to
resolve strong discontinuities with a few grid points and without
producing damping of the solution, and the energy is well
conserved. The PML boundary condition allows us to calculate long
temporal series of simulations without reflections of the waves
reaching the top boundary.

We have presented the analysis of several simulations where the
sunspot atmosphere was perturbed by different pulses, varying
their location and temporal behaviour. The simulations of short
period waves in the three-dimensional sunspot model clearly show
several phenomena that are predicted by wave theory. We confirmed
that our code correctly describes the propagation of slow and fast
modes, both in regions dominated by the magnetic field or the gas
pressure. Our main findings, learned from the simulations, can be
summarized as follows:
\begin{itemize}

\item The conversion between fast and slow magneto-acoustic waves
happens in three dimensions in a qualitatively similar way as in
two dimensions. Waves with frequencies down to the cut-off
frequency behave in the same way. The driver located in the gas
pressure dominated region generates mostly the fast (acoustic)
mode. This mode, propagating to the upper layers, is transformed
at the height where $c_S=v_A$. After the transformation, a slow
acoustic mode propagates upwards along the field lines in the
magnetically dominated atmosphere. The fast magnetic mode
undergoes refraction and it is reflected back to the
sub-photosphere. When it reaches again the surface $c_S=v_A$, new
transformations take place producing another fast acoustic and
slow magnetic modes in the region $v_A < c_S$.

\item High-frequency field-aligned propagating acoustic waves are
constantly produced in the upper magnetically dominated atmosphere
at locations away from the source. These waves appear due to the
continuous transformation from the fast (acoustic) waves moving
horizontally, across the field lines, away from the source in the
gas pressure dominated region. On their way, the fast waves
constantly touch the $c_S=v_A$ layer producing slow (acoustic)
waves in the upper atmosphere in the horizontal locations far
from the source. We observe this behaviour in all simulations with
different driving frequencies and source position.

\item The 3D simulations allow us to identify an Alfv\'en mode.
This mode appears only in the simulation with the source located
away from the sunspot axis. It is produced after the
transformation from the fast (acoustic) mode. We find that the
transformation efficiency from the fast to the Alfv\'en mode is much
lower than that from the fast to the slow mode. The eventual energy of the
Alfv\'en wave in the magnetically dominated region is 20 times
lower than that of the slow mode. In the simulations with the
driver located at the axis of the sunspot, where the angle between
the direction of the upwards propagating wave and the magnetic
field is zero, we find no indications of the transformation to the
Alfv\'en mode.

\item The analysis of the wave energy fluxes suggests that in the
high-frequency cases (above the cut-off) the wave energy can reach
the upper atmosphere most efficiently in the form of slow
(acoustic) field aligned propagating waves. After some height in
the middle photosphere there is no magnetic flux corresponding to
the fast (magnetic) waves as their energy is reflected. If the
driver is located away from the axis, some small part of the energy
also can propagate upwards in the form of an Alfv\'en wave.

\item Both magnetic and acoustic energy of the low-frequency waves
(smaller than the cut-off) remain almost completely below the
level $c_S=v_A$ and do not reach upper layers. This happens
because the energy of the fast magnetic modes in the upper layers
is reflected, and the evanescent acoustic slow modes do not
propagate any energy at these frequencies. The comparison between
the magnetic fluxes at high and low frequencies shows that the
magnetic flux reaches smaller heights for the low-frequency waves.

\item When the driver excites a spectrum of waves, we observe a
change of the dominant frequency of oscillations with height from
the photosphere to the chromosphere. The driver excites a spectrum
with a maximum power at 3.33 mHz frequency. Waves at this frequency
are evanescent in the atmosphere. At higher frequencies (above
$\nu_c=5.8$ mHz in our sunspot model) the waves can propagate
upwards along the field lines.  Due to the larger amplitude
increase with height of the propagating waves, compared to the
evanescent waves, the 3-minute waves ($\nu \approx 5.9$ mHz)
dominate the power spectrum in the chromosphere. This behavior,
obtained in the simulations, is similar to the observed one.
\end{itemize}

One of the questions that arises from these results is the
evaluation of the validity of the decomposition performed to
separate the fast mode from the Alfv\'en mode in the low-beta
plasma. The decoupling following \citet{Cally+Goossens2008} is
valid for an idealized case when considering a uniform magnetic
field, for a plane wave with a constant wavenumber perpendicular
to gravity, and it is obtained asymptotically in the limit of
infinite Alfv\'en speed. Although the realistic atmosphere used in
our calculations does not fulfill these restrictions, a coherent
picture is retrieved from the magnetic flux of the velocity and magnetic
field components in the direction given by Eq. \ref{eq:eperp},
showing upwards propagation along magnetic field lines. The result that
naturally emerges from this decomposition justifies the application of the method.

Another issue that complicates the decoupling of both modes is the
excitation of small horizontal wavenumbers due to the limited
horizontal extent of the driver. The fast mode is refracted due to
the rapid increase of the Alfv\'en speed with height, but the
altitude at which it happens depends on the wavenumber, being
higher for lower wavenumbers. Since most of the power excited by
our driver lies in the range of wavenumbers below $\sim
1/R_{scr}$, some of the fast mode waves might still be
partially refracted in the limited 1 Mm atmosphere above the $c_S=v_A$
surface. The magnetic flux of these waves may reach high layers,
complicating the separation of the fast and Alfv\'en modes. In this scenario, we would expect a continuous
transition between the upward longitudinal magnetic flux and the
refracted one at different heights where all these waves are being
refracted. However, the magnetic flux of the Alfv\'en wave that
we retrieve is clearly delimited along magnetic field lines,
confirming that this flux mostly corresponds to the Alfv\'en
mode.

The most important achievement reached by the development of our
numerical code is the possibility to investigate the
three-dimensional mode transformation in realistic conditions
imitating a sunspot atmosphere. This, together with the
possibility to study large-period waves in the layers where they
are observed, gives an opportunity for the direct comparison
between our numerical simulations and solar spectropolarimetric
observations. Simulations of sunspot high layers represent a hard
challenge due to the exponential increase of the Alfv\'en speed
with height. Our sunspot model presents an Alfv\'en speed of
almost 1000 \kms\ close to the upper boundary of the domain,
limiting the time step and making the calculations very expensive.
Despite this, our code manages to describe the waves well,
including the correct performance of the boundary PML layer. Note
that other works on waves in non-trivial magnetic configurations
have been restricted either to two-dimensional high-frequency
cases \citep{Cargill+etal1997, Rosenthal+etal2002, Hasan+etal2003,
Bogdan+etal2003, Khomenko+Collados2006} or to the study of
helioseismic waves, where the problem of high Alfv\'en speed is
avoided \citep{Cally+Bogdan1997, Parchevsky+Kosovichev2009,
Hanasoge2008, Cameron+etal2008, Moradi+etal2009,
Khomenko+etal2009}.

The strategy applied in our code allows the direct comparison with
observations by means of spectral synthesis. The simulations
presented in the paper reproduce the frequency change with height,
indeed observed in the sunspot atmospheres
\citep{Centeno+etal2006a}. In the future we plan to perform a more
detailed comparison with solar data. This will be done by
exciting the sunspot magneto-static model in equilibrium with
velocities obtained from spectropolarimetric observations, and by
comparison of the simulated wave parameters in the photosphere and
chromosphere with those obtained from simultaneous observations in
different spectral lines.

Perhaps the most interesting simulation considered in our paper is
the one with the source exciting a spectrum of waves close to the
solar one. This case has a special relevance  because theoretical
models of wave transformation, existing as of today, are best
valid in the high-frequency limit and do not address the behavior
of waves at frequencies below the cut-off frequency
\citep{Schunker+Cally2006}. Our initial results show that almost
no energy of the 5-minute waves propagates into the higher layers,
at least in the situation of the sunspot model and source located
at the axis considered in the paper.
These results need further exploration. In our future work we will
perform simulations of 5-minute waves in larger
computational boxes, varying the excitation mechanisms (location
and properties of the source), as well as the sunspot model.
In particular, an interesting question is under which conditions
5-minute Alfv\'en waves can be excited by the mode
transformation. These waves can still propagate some energy into
the upper atmosphere of solar active regions, and thus
understanding the conditions of transformation to these waves and
their energetics in sunspots is important.
The number of works in the literature with numerical calculations
including the transformation to Alfv\'en waves is scarce. The most
relevant study is the one by \citet{Cally+Goossens2008} who find
that the transformation to an Alfv\'en mode is effective at certain
angles of inclination and azimuth of the magnetic field.

\acknowledgements  This research has been funded by the Spanish
MICINN through projects AYA2007-63881 and AYA2007-66502. The
simulations have been done on LaPalma supercomputer at Centro de
Astrof\'{\i}sica de La Palma and on MareNostrum supercomputer at
Barcelona Supercomputing Center (the nodes of Spanish National
Supercomputing Center). To represent the data we have used VAPOR
software (http://www.vapor.ucar.edu).

%\aareferences

\appendix

\section{Tests on numerical performance}
\label{sect:test}

In this appendix we describe the results of standard numerical
tests to verify the code performance.

\subsection{1D Riemann shock tube test}
The Riemann shock tube test \citep{Sod1978,Caunt+Korpi2001} has
been simulated in order to test the behavior of the
hydrodynamical part of the code, including discontinuities in the
properties of the fluid. This allow us to analyze how the
artificial viscosity copes with shock capturing. The physical
domain size is unity and the initial conditions include a
discontinuity at $x=0.5$. On the left we have density $\rho_1=1$
and pressure $p_1=1$, while on the right $\rho_1=0.125$ and
$p_1=0.1$. The ratio of specific heats is $\gamma=1.4$ and the
initial velocity as well as the magnetic field are set to zero.
The problem has been simulated in one dimension with a resolution
of 256 grid points and closed boundaries.

\begin{figure}
\centering
\includegraphics[width=9cm]{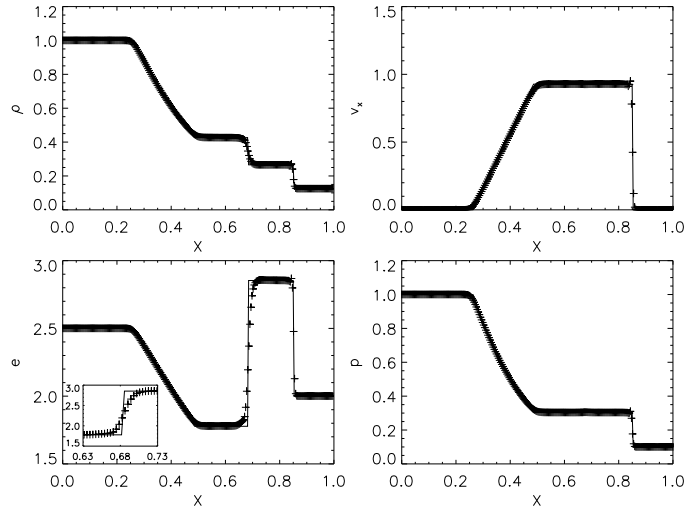}
\caption{Standard Riemann shock tube problem at $t=0.2$. Crosses
represent numerical solution, while lines represent the analytic
solution with an exact nonlinear Riemann solver. Inner plot on
left bottom panel helps to visualize better the discontinuity at
$x=0.68$.} \label{fig:riemann}
\end{figure}

Figure \ref{fig:riemann} shows the density, velocity, internal
energy density per unit mass and pressure at time t=0.2 for the
simulation compared to the analytical solution. From left to right
the plot shows a rarefaction wave (from $x=0.25$ to $x=0.5$), a
contact discontinuity (at $x=0.68$) and a shock front ($x=0.85$).
The position of all of them matches precisely with the analytical
solution and the magnitudes of the fluid properties are correct.
The contact discontinuity for the energy was inevitably smoothed,
as shown in the plot at the bottom left where a region around it
is amplified, but still a 93\% of the amplitude of the
discontinuity is covered with 10 grid points. Moreover, the shock
front is resolved with 3 grid points, which proves the good
performance of the code in shock capturing.

\subsection{1.5D Brio \& Wu shock tube}
\label{sect:brio&wu} To test the formation of magnetohydrodynamic
shock waves we use the MHD analog of the Sod shock tube problem
described by \citet{Brio+Wu1988}, which has been widely used in
previous works \citep{Stone+Norman1992b, Caunt+Korpi2001,
Shelyag+etal2008}. We can compare our results with those given in
the literature, since no known analytical solution exists for the
evolution of this problem. In this 2D test the fluid is
initialized in a physical domain from $x=0$ to $x=1$ and with a
discontinuity in density, pressure and magnetic field normal to
the direction of motion located at $x=0.5$. Parameters at the left
hand side from the discontinuity are $\rho_1=1$, $p_1=1$ and
$B_{y1}=\sqrt{\mu_0}$, and at the right hand side are
$\rho_2=0.1$, $p_2=0.1$ and $B_{y2}=-\sqrt{\mu_0}$. All the domain
is permeated with a constant magnetic field along the direction of
motion $B_x=0.75\sqrt{\mu_0}$ and the adiabatic index $\gamma$ is
set to 2. In this case the resolution is 800 grid points in the
direction of the shock wave propagation, similar to the other
published works. All the boundaries are closed.

\begin{figure}
\centering
\includegraphics[width=9cm]{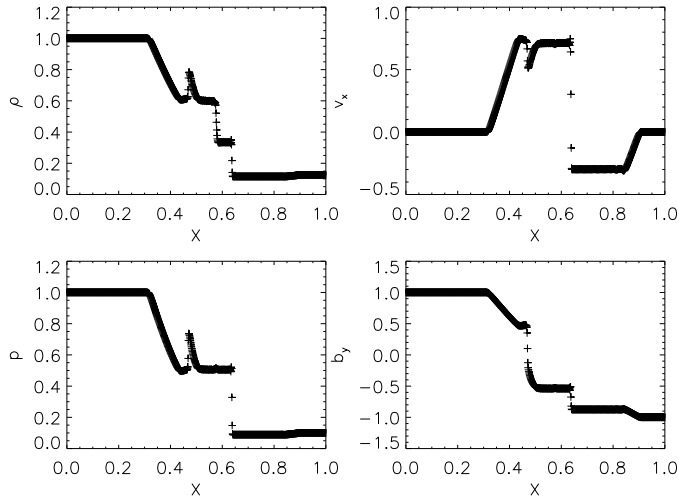}
\caption{Numerical solution of the Brio \& Wu problem at t=0.11
(see \S\ \ref{sect:brio&wu}). } \label{fig:brio}
\end{figure}

The density, velocity in $x$ direction, pressure and magnetic
field in $y$ direction are shown in Figure \ref{fig:brio}. This
MHD Riemann problem produces a complex solution with several
components: the waves moving to the left are a fast rarefaction
wave and a slow compound wave (consisting of a slow rarefaction
attached to a slow shock), and the waves moving to the right
include a contact discontinuity, a slow shock and a fast
rarefaction wave. Comparison of our results with other works show
that the waves have propagated with the correct velocities and
have similar magnitudes, indicating that they are in good
agreement with other solutions. The slow shock is resolved again
with only 3 grid points.

\subsection{2D Orsz\'ag-Tang vortex}
\begin{figure}
\centering
\includegraphics[width=9cm]{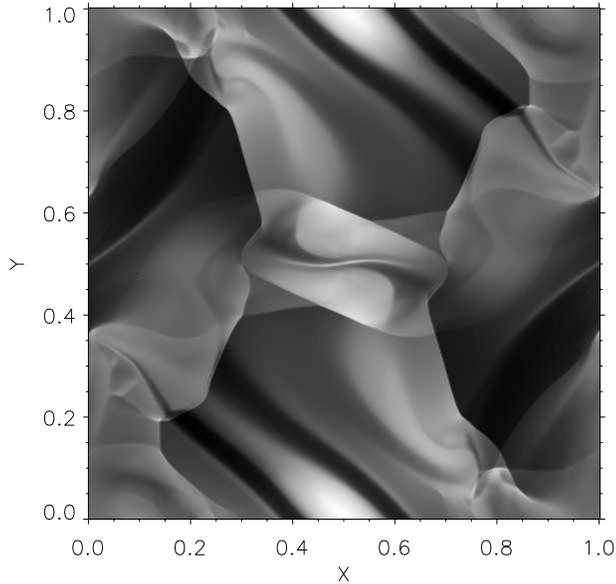}
\caption{Mass density at t=0.5 of the Orsz\'ag-Tang vortex
simulation.} \label{fig:orszagtang}
\end{figure}

The next test is the Orszag-Tang vortex, which was originally
studied by \citet{Orszag+Tang1979} and has been used to probe
several codes \citep{Ryu+etal1995, Dai+Woodward1998,
Londrillo+delZanna2000, Shelyag+etal2008}. This problem allows us
to demonstrate the robustness of the numerical scheme used in our
code solving the two-dimensional interaction of non-linear
shock-waves and also to compare qualitatively the code with other
codes. The initial conditions for density and gas pressure are
constant, with $\rho=25/(36\pi)$ and $p=5/(12\pi)$, the magnetic
field $B_x=-\sin(2\pi y)$ and $B_y=\sin(4\pi x)$ and the initial
velocity $v_x=-\sin(2\pi y)$ and $v_y=\sin(2\pi x)$. Therefore,
the initial flow is a velocity vortex superimposed to a magnetic
vortex, with a common X-point, but with different structure. The
initial Mach number is $M_0=1$ and the adiabatic index is set to
$\gamma=5/3$. In our simulation for this problem we have chosen a
unit size in horizontal and vertical dimensions and the resolution
of the simulation box is set to \hbox{512 $\times$ 512} grid
points. Figure \ref{fig:orszagtang} presents the density at time
$t=0.5$, showing precise agreement with the other published works.

\subsection{3D Acoustic wave}

\begin{figure}
\centering
\includegraphics[width=9cm]{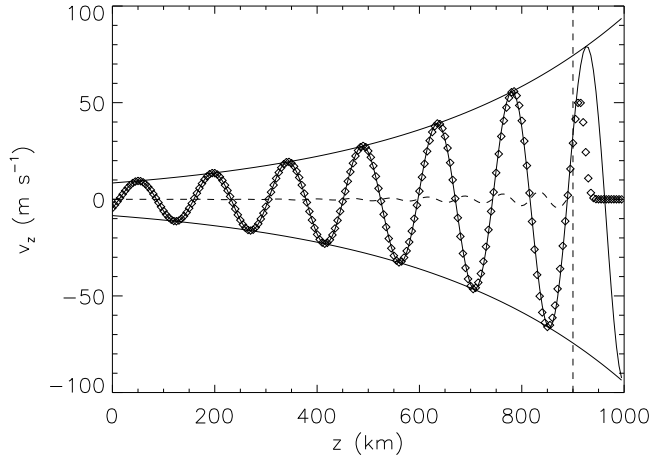}
\caption{Vertical velocity profile of vertically propagating
acoustic waves in an isothermal, stratified atmosphere with
vertical constant magnetic field at t=197 s. Solid line: exact
solution; diamonds: numerical solution. Dashed line is the
difference between both solutions. The vertical dashed line
indicates the location of the PML interface.} \label{fig:acoustic}
\end{figure}

Since our code is oriented for the simulations of waves, it is
necessary to test how well it can approximate the known analytical
solutions for different types of waves in a stratified atmosphere,
their propagation speeds, amplitudes and shock development. The
analytical solution of an acoustic wave propagating in an
isothermal atmosphere with vertical stratification due to gravity
and permeated with a constant vertical magnetic field is known
from \citet{Ferraro+Plumpton1958}. The vertical velocity for wave
with frequencies $\omega$ above the cut-off frequency
$\omega_c=\gamma g/(2c_S)$ follows as:
\begin{center}
\begin{equation}
\label{eq:acoustic} v_z(z,t)=De^{z/2H_0}\sin{\Big
[(\frac{\sqrt{\omega^2-\omega_c^2}\big)}{c_S}z + \omega t \Big ]},
\end{equation}
\end{center}
where $H_0=c_S^2/(\gamma g)$ is the pressure scale height and
$c_S$ is the sound speed. This analytical solution with a period
of 15 s and starting amplitude of 10 \ms\ has been introduced as a
bottom boundary condition without variations in the horizontal
directions and its evolution in time has been calculated with the
numerical code. The physical domain is set to $1000$ km in the
vertical direction with a resolution of 200 grid points, using 20
of them as PML domain at the top. In both horizontal dimensions
the computational domain consists of 7 grid points, covering $35$
km, and periodic boundary conditions were used in these
directions. In Fig. \ref{fig:acoustic} we show the comparison
between the numerical and the analytical linear solution for the
vertical velocity after 197 s of simulations. The numerical
solution matches the exact solution in the computational domain,
while it is damped effectively by the PML layer. It is important
to note that the amplitude and propagation speeds are both
described correctly by the numerical solution and that the
numerical diffusion does not affect the amplitude increase with
height. No spurious reflections are present showing the correct
action of the PML boundary in this example.

The dashed line represents the difference between both solutions,
which also increases with height. The main contribution to this
difference are the nonlinear terms that are taken into account in
the numerical calculation, but not in the analytical solution. We
have checked that reducing the driver amplitude by a factor of two
reduces the difference by exactly a factor of four. The wavelength
of the difference is twice shorter compared to that of the wave.
These two arguments prove that the difference between the
analytical and numerical solutions is mostly due to non-linear
terms.

\subsection{3D Alfv\'en wave}
As a next step, the response of the numerical scheme to the
propagation of an Alfv\'en wave in an isothermal, stratified
atmosphere with vertical magnetic field is analyzed. The
analytical solution was developed by \citet{Ferraro+Plumpton1958},
and, according to \citet{Khomenko+Collados+BellotRubio2003}, the
solution for the horizontal velocity can be written as

\begin{figure}
\centering
\includegraphics[width=9cm]{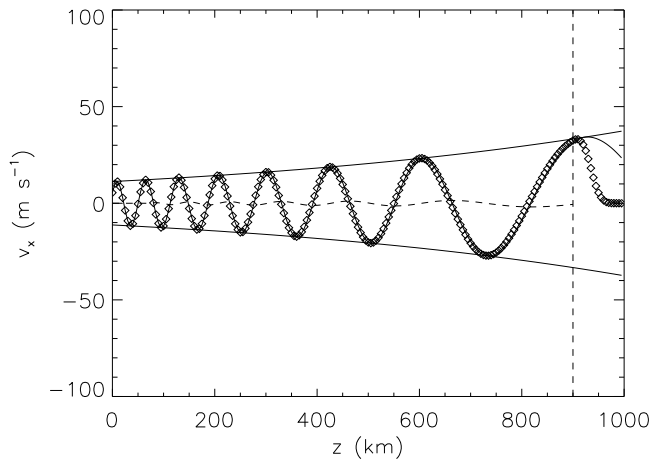}
\caption{Horizontal velocity profile of Alfv\'en waves in an
isothermal, stratified atmosphere with vertical constant magnetic
field at t=115 s. Solid line: exact solution; diamonds: numerical
solution. Dashed line is the difference between both solutions.
The vertical dashed line indicates the position of the PML
interface.} \label{fig:alfven}
\end{figure}

\begin{center}
\begin{equation}
\label{eq:alfven} v_y(z,t)=i\xi_0\omega \sqrt{J^2_0+Y^2_0}{\rm
exp}\Big[i\Big(\omega t +{\rm arctang} \frac{Y_0}{J_0}\Big )\Big]
\end{equation}
\end{center}

\noindent where $J_0$ and $Y_0$ are Bessel functions and $v_A$ is
the Alfv\'en speed. The same atmosphere as in the previous section
was used in this test, but now the bottom layers were excited with
the solution of an Alfv\'en wave of period 10 s and an amplitude
of 10 m s$^{-1}$ as a boundary condition. This driver generates
the propagation of Alfv\'en waves toward higher layers of the
atmosphere. The horizontal velocity of both the numerical and
exact solutions for time \hbox{$t=115$} $\rm{s}$ is shown in Fig.
\ref{fig:alfven}, demonstrating a very good match.

The dashed line shows the difference between the numerical and the
analytical solution. In this case, the nonlinearities are not so
important since the amplitude is lower than the one of the
acoustic wave, and the discrepancy between both solutions is due
to the reflection produced at the top boundary. The PML layer
results more problematic for transversal waves which oscillate
parallel to the interface between the PML media and the physical
domain, and they give rise to reflections of 5-6\% of the
velocity.

\subsection{3D Strong blast wave}

Our last test consists of the explosion of a spherical high gas
pressure region in a magnetized, initially static 3D medium. It
has been commonly used for code validation \citep[see, for
example][]{Balsara+Spicer1999,Londrillo+delZanna2000}, and the
set-up consists of a cubic domain with 256 grid points in the
three spatial dimensions spanning from 0 to 1. The initial
density, $\rho_0$, is set to unity in all the domain, while the
initial pressure is set to unity all over except a spheric hot gas
region located at the center of the domain of radius $r_0=0.125$,
which is a hundred times overpressured ($p_1=100$). A constant
magnetic field with a strength of $B_0x=10\sqrt{\mu_0}$ is
initialized along the $x-$direction.

\begin{figure}
\centering
\includegraphics[width=9cm]{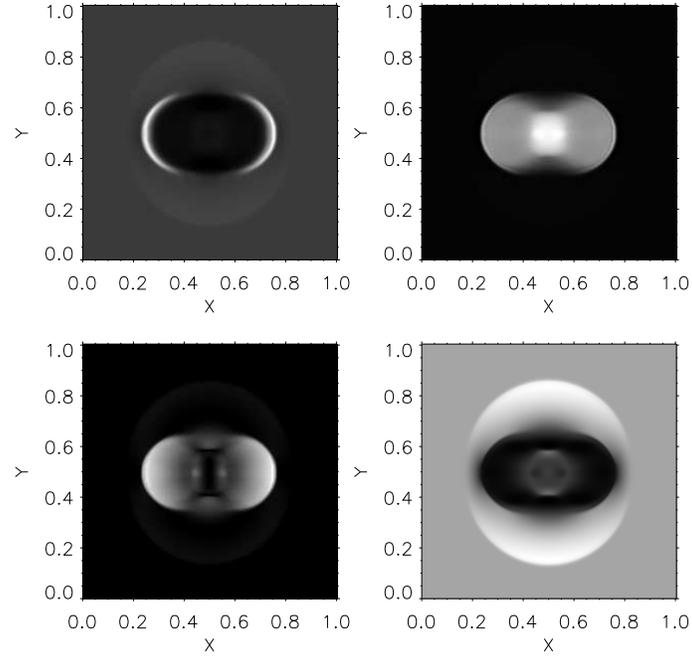}
\caption{Numerical solution of the 3D strong blast wave density at
the plane $x-y$ and $z=0.5$ at $t=0.02$. From left to right and
from top to bottom: Mass density, pressure, velocity squared and
magnetic pressure.} \label{fig:blast}
\end{figure}

In Fig. \ref{fig:blast} we show a cut of the density in the plane
$x-y$ at $z=0.5$ and $t=0.02$. The system shows the axial symmetry
imposed by the magnetic field. We can identify the different wave
modes present in the simulations. The outermost wave corresponds
to the fast magnetoacoustic mode, and inside this region there are
two wave fronts propagating along the magnetic field, which is a
slow magnetoacoustic shock. This test verifies that our code can
handle with three dimensional propagation of highly nonlinear
waves at the correct propagation speed, resolving the shocks
thanks to the hyperdiffusive terms.

\end{document}